\documentclass[
 reprint,
groupedaddress,
amsmath,amssymb, 
aps,
pre,
]{revtex4-2}

\usepackage{ulem}
\usepackage[english]{babel}
\usepackage{dsfont}
\usepackage{amsmath}
\usepackage{hyperref}
\usepackage{mathrsfs}
\usepackage{bm}
\usepackage{cancel}
\usepackage{graphicx}%
\usepackage{dcolumn}
\usepackage{amsmath}

\usepackage{placeins}
\usepackage{comment}
\usepackage[bottom]{footmisc}
\usepackage{xcolor}
\usepackage{physics}
\usepackage{listings}
\usepackage{bbold}
\newcommand{\RN}[1]{%
  \textup{\uppercase\expandafter{\romannumeral#1}}%
}

\newcolumntype{C}{>{$}c<{$}}
\AtBeginDocument{
\heavyrulewidth=.08em
\lightrulewidth=.05em
\cmidrulewidth=.03em
\belowrulesep=.65ex
\belowbottomsep=0pt
\aboverulesep=.4ex
\abovetopsep=0pt
\cmidrulesep=\doublerulesep
\cmidrulekern=.5em
\defaultaddspace=.5em
}

\usepackage{floatrow}

\usepackage{booktabs}

\usepackage[labelformat=simple,subrefformat=parens,labelformat=parens]{subfig}
\newcommand{\mat}[1]{\underline{\underline{\bm{#1}}}}
\usepackage{caption}
\definecolor{sclr}{rgb}{0.27, 0.51, 0.71} 
\definecolor{BrighterTeal}{HTML}{005F56}
\definecolor{RichPlum}{HTML}{5B2A3F}
\definecolor{DeepBlue}{HTML}{2F4A72}
\definecolor{WarmOchre}{HTML}{A36D3E}

\begin{document}

\title{Harvesting chemical power from cyclic environments}

\author{Pranay Jaiswal, Ivar S. Haugerud, Hidde D. Vuijk, Christoph A. Weber}
\email[]{christoph.weber@physik.uni-augsburg.de}

\affiliation{Faculty of Mathematics, Natural Sciences, and Engineering: Institute of Physics, University of Augsburg, Universit\"atsstra{\ss}e\ 1, 86159 Augsburg, Germany}

\date{\today}

\begin{abstract}
Life relies on a sophisticated metabolic molecular machinery
that turns over high-energy molecules to evolve complex macromolecules and assemblies. 
At the molecular origin of life, such machinery was absent, implying the need for simple yet robust physical mechanisms to harvest energy from the environment and perform chemical work or produce chemical power. However, the mechanisms involved in harvesting energy from a macroscopic cyclic environment to drive chemical processes on the molecular scale remain elusive. 
In this work,  we propose a theory that describes the kinetics of chemical reactions in a system subject to a cyclic reservoir with varying properties. We compare cycles of solvent (wet-dry cycles), with cycles of a component participating in a chemical reaction (reactant cycle). We find that for both wet-dry and reactant cycles, resonance frequencies exist at which the chemical power is maximal.  We identify which cycle type is more beneficial in harvesting chemical power for different molecular interactions. 
Our findings of harvest efficiencies around ten percent suggest that the cyclic environment could have played a key role in catalyzing the metabolic molecular machinery at the molecular origin of life. 
\end{abstract}
\maketitle

\begin{figure*}[tb]
    \centering
     \makebox[\textwidth]{\includegraphics[width=\textwidth]{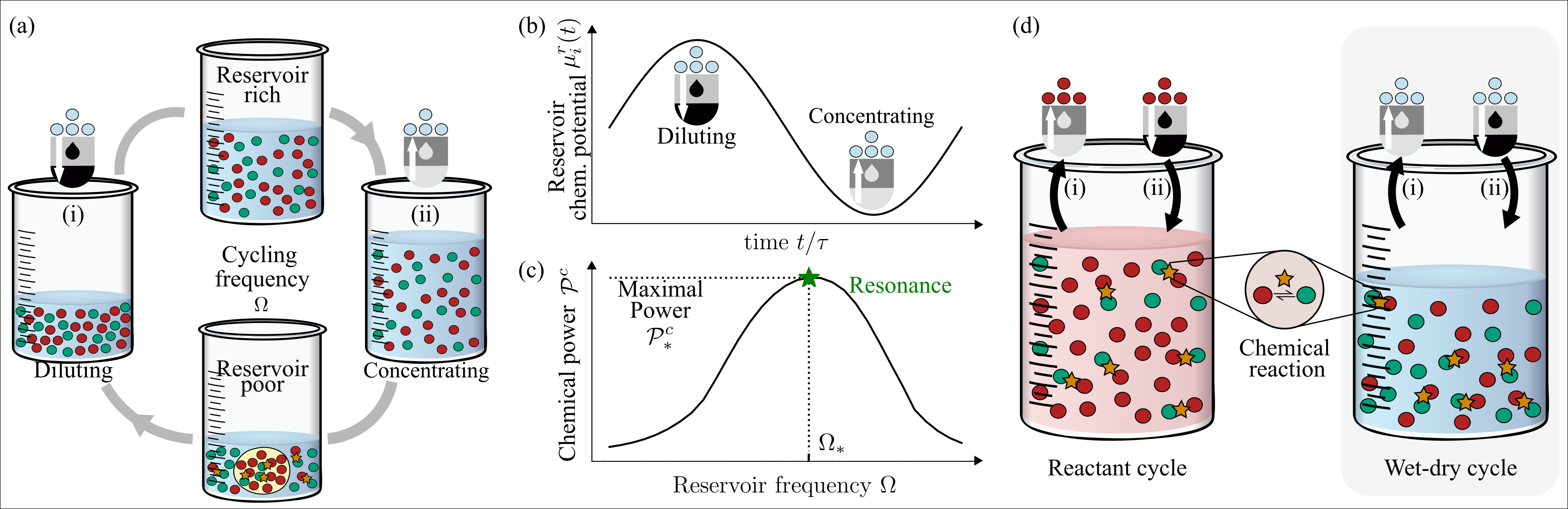}}
     \caption{\textbf{Harvesting chemical power from a cyclic reservoir.} 
     (a) A mixture with a chemical reaction is coupled to a cyclic reservoir. The reservoir periodically exchanges solvent (wet-dry cycles) or reactant (reactant cycles).
     Such components are either added (ii) or removed (i) oscillating between a state rich or poor in the component exchanged with the reservoir (``reservoir-rich/poor''). 
     In certain conditions, the system can reach a non-dilute regime, where interactions among the components lead to liquid-liquid phase separation.
     (b) The cyclic property of the reservoir is governed by an oscillating chemical potential of the exchanging component.
     (c) We calculate the chemical power and find that there is a resonance frequency $\Omega_*$ at which the chemical power is maximal.
     (d) The value of the resonance frequency and maximal chemical power vary depending on the cycle type (wet-dry cycle or reactant cycle).}
     \label{fig:systems_illistration}
\end{figure*} 
\section{Introduction}
\label{Introduction}
Living systems turnover energy and matter through which they  maintain themselves away from thermodynamic equilibrium~\cite{phillips2012physical}. Specifically, they developed a metabolic machinery that involves
complex macromolecules~\cite{alberts2015essential}, in particular proteins and protein assemblies
that can hydrolyze highly energetic ``fuel'' molecules, such as ATP or GTP.
Such macromolecules include, as examples, ribosomes used to synthesize proteins~\cite{bassler2019eukaryotic}, or  molecular motors capable of performing mechanical work~\cite{howard2002mechanics, schliwa2003molecular}, among many others.
This metabolic molecular machinery, which utilizes fuel molecules to evolve the system's complexity, is a hallmark feature of life and defines it as being out of equilibrium or as ``active matter"~\cite{gentile2023active}.

At the molecular Origin of Life,  none of such molecular machinery was available -- it had to evolve through selection along with the emergence of life itself~\cite{sutherland2017opinion,kriebisch2025roadmap}. The synthesis of such a machinery required a mechanism to harvest energy from the surroundings making it available for chemical processes on the molecular scale -- 
an aspect that is not only relevant for the molecular Origin of Life, but also for the emergence of \textit{de novo} life~\cite{adamski2020self, kriebischRoadmapSynthesisLife2025}. 
The synthesis of such machinery is unlikely close to thermodynamic equilibrium due to its higher free energy compared to the machinery's building blocks. Thus,  non-equilibrium conditions had to act similar to a metabolism.
Such conditions can be  created by a  surrounding with physical properties (salt, temperature, composition, etc.) varying in time~\cite{otto2021approach, ianeselli2023physical,haugerud2024nonequilibrium}. For a sustainable evolution, these time variations have to occur in cycles, or in some thought of periodic variations, often involving a characteristic average frequency.  

Classical examples relevant to the molecular origin of life are day-night cycles that lead to periodic changes  in temperature and UV light~\cite{westall2018hydrothermal, sasselov2020origin}. 
Cyclic non-equilibrium conditions can also occur due to faster weather changes affecting temperature, humidity, or pressure~\citep{mulkidjanian_origin_2012,damer_coupled_2015}, or give rise to salt deliquescence~\citep{campbell_prebiotic_2019}, or freeze-thaw cycles~\citep{vlassov_ligation_2004, le_vay_enhanced_2021,mutschler_freezethaw_2015}. 
The cycles mentioned take place on time-scales between minutes to roughly a day, with a characteristic frequency range from  $10^{-2} \, \text{s}^{-1}$  down to  $ 10^{-5} \, \text{s}^{-1}$. 
Moreover, even slower cycles of the system's solvent content can be found in foams~\cite{tekin_prebiotic_2022}, or porous rock containing trapped gas-bubbles~\cite{ianeselli_non-equilibrium_2022,matreux_heat_2021}.
Similarly, wet-dry  or geological cycles cause oscillatory changes in solvent or reactive components~\cite{becker2018wet, damer2020hot}. 
A recent specific example of wet-dry cycles affecting chemical processes is the evolution of long DNA strands
in a Hadean CO2 atmosphere~\cite{ianeselli2022water}. 
Extreme cycle conditions can further lead to non-dilute conditions and phase-separation of the liquid mixture~\cite{fares2020impact, bartolucci2023sequence}. However, 
up to now, the general mechanism underlying the harvest of chemical power from cyclic non-equilibrium conditions remains elusive, and it is unclear under what conditions energy is efficiently transferred from the macroscale to the scale of chemical reactions.

In this work, we develop a theory for liquid mixtures of chemically reacting components coupled to a cycling reservoir, as depicted in Fig.~\ref{fig:systems_illistration}. Specifically, we consider reservoirs that can exchange solvent molecules with the system, generating wet-dry cycles, or a reservoir that cycles the reacting components of the chemical reaction, leading to reactant cycles. Our key finding reveals that chemical power can be harvested by driving the system through reservoir cycles. We found that there is a resonance frequency at which cycling both reservoirs maximizes the chemical power. The resonance phenomenon can be understood through an analogy of coupled harmonic springs driven by the cycling reservoir. Depending on the rate of exchange with the reservoir, the harvested power can be greater for wet-dry or reactant cycles. We further elucidate how liquid-liquid phase separation influences maximal chemical work and the resonance frequency. Strikingly, the harvest at resonance is characterized by high efficiencies, where around 10\% of the reservoir's dissipated energy can be harvested from the reservoir and transferred to the molecular level as chemical work or power. 
Our results suggest that a cycling reservoir can provide conditions similar to a low-efficiency catalyst that can kick off chemical processes forming high-energy complex molecules at the molecular origin of life -- an essential step for the emergence of a prebiotic metabolism.

\section{Theory for harvesting chemical power from a cyclic environment}

In this section, we propose a 
theory for a mixture with one representative chemical reaction. 
The components undergoing chemical reactions in the mixture are non-dilute, implying that interactions among the components are affecting the diffusion and reaction fluxes. 
The mixture is coupled to a reservoir that can exchange components with the mixture. 
The reservoir chemical potentials undergo cycles, mimicking a cyclic environment that maintains the mixture out-of-equilibrium. 

We first present the continuum theory that describes the coupling to the cyclic reservoir through the boundary conditions (Sect.~\eqref{eq:cont_theory}).
Second, we derive a significantly reduced theory valid for a system small enough such that concentration fields are approximately homogeneous in each phase (Sect.~\eqref{eq:theory_phaseeq}). 
The corresponding solutions approach the ones from the continuum theory when the largest reaction-diffusion length scale increases beyond the system size. 
In our work, we ask how much chemical work and power can be harvested compared to the energy dissipated by cycling the reservoir (Sect.~\ref{sect:chem_work}). 
The reduced theoretical approach enables us to understand the conditions of optimal energy harvest through a mapping to a two-coupled and driven harmonic oscillators elongating and retracting in the mixture's thermodynamic phase diagram.    

\subsection{Continuum description of a reacting mixture subject to cycling reservoirs}\label{eq:cont_theory}

The continuum theory for the mixture is comprised of the spatial and temporal evolution of the volume fraction fields $\phi_i(\bm{x}, t)$. 
These fields are governed by diffusive fluxes $\bm{j}_i$ and the net source $s_i$~\cite{julicherDropletPhysicsIntracellular2024,haugerud2025excitability, bauermannEnergyMatterSupply2022d, weberPhysicsActiveEmulsions2019d}:
\begin{equation}   \partial_t\phi_i(\bm{x}, t) =  -\nabla \cdot \bm{j}_i(\bm{x}, t) + s_i(\bm{x}, t)\,  ,
\label{eq:phi_dot}
\end{equation}
where spatial gradients in chemical potentials, $\nabla \mu_i$, drive diffusive fluxes, $\bm{j}_i = -\Lambda_i \nabla \mu_i$, with $\Lambda_i$ denoting the mobility coefficients~\cite{de2013non}. 
The chemical potential $\mu_i=\nu_i \delta G/\delta \phi_i$ can be derived from the mixture's Gibbs free energy $G$; see Appendix~\ref{app:gibbs_free_energy} for the free energy and derivation of chemical potential.\par

The surface boundary of the system consists of two parts: $\partial \Theta_r$ which is the surface area in contact with the reservoir, and $\partial \Theta$, which is an impenetrable surface boundary and therefore has a no-flux boundary condition $\bm{\hat{n}}\cdot\nabla\mu\rvert_{\partial\Theta}=0$, where $\hat{\bm{n}}$ denotes the surface normal vector.
When the mixture's surface is in contact with the reservoir through the surface $\partial \Theta_r$, there is a reservoir flux $\tilde{\bm{h}}_i$ through the surface element $\text{d}^2\bm{x}$ normal to $\partial \Theta_r$ (with the unit of a rate per area).
This flux is governed by the chemical potential of the bulk $\mu_i$ and the reservoir $\mu_i^r$:
\begin{equation}
    \tilde{\bm{h}}_i(\bm{x}_b) = \hat{\bm{n}} \, \tilde{k}_{r,i} \left[\exp\left\{\frac{\mu_i^r}{k_\text{B}T}\right\}-\exp\left\{\frac{\mu_i(\bm{x}_b)}{k_\text{B}T}\right\}\right]\,,\label{eq:boundary_flux}
\end{equation}
where $\tilde{k}_{r,i}$ is the reservoir flux rate coefficient with the unit of a rate per area. 
Moreover,  $k_\text{B}T$ is the thermal energy, and $\bm{x}_b$ is a point on the mixture's boundary that is in contact with the reservoir, $\bm{x}_b \in \partial \Theta_r$.
Local equilibrium with the reservoir is reached when chemical potentials are equal, $\mu_i(\bm{x}_b)=\mu_i^r$ for all points $\bm{x}_b$ at the mixture's boundary $\partial \Theta_r$, leading to a vanishing flux $\tilde{\bm{h}}_i(\bm{x}_b)=0$.

The exchange of molecules between the mixture and the reservoir affects the mixture's composition and volume. The boundary moves with a speed
\begin{equation}
    v(\bm{x}_b, t) = \sum_i \nu_i\tilde{\bm{h}}_i(\bm{x}_b)\cdot\bm{\hat{n}}\,,
\end{equation}
which further alters the system volume $V$ according to
\begin{equation}
    \dv{V}{t} = \int_{\partial\Theta_r}\text{d}^2\bm{x}\,\, v(\bm{x}_b, t)\,.
\end{equation}

In this work, we describe the cyclic environment by a periodically cycling reservoir chemical potential $ \mu_i^r(t)$ as shown in Fig.~\ref{fig:systems_illistration}(b).
To model a characteristic frequency of the various cycle types mentioned in the introduction, we consider a sinusoidal form with a frequency $\Omega$ and an amplitude \(\hat\mu_i\) around an average reservoir value \(\left<\mu_i^r\right>\):
\begin{equation}
    \mu_i^r(t) = \left<\mu_i^r\right> + \hat\mu_i \sin\left(2\pi \Omega \, t\right) \, . 
    \label{rev_osi}
\end{equation}
As illustrated in Fig.~\ref{fig:systems_illistration}(d), we consider two  distinct cases of the reservoir cycle:
\begin{itemize}
\item [(i)] Periodic exchange of the solvent between the reservoir and the mixture, called \textit{wet-dry cycles}.   In this case, the reservoir exchange rate coefficient has the form $\tilde{k}_{r,i}=k_{r}\delta_{i,S}$ with $S$ denoting the solvent component.
\item[(ii)] Periodic exchange of a reacting component  between the reservoir and the mixture, referred to as \textit{reactant cycles}. For the case of changing the reacting component $A$,  the reservoir exchange rate coefficient has the form $\tilde{k}_{r,i}=k_{r}\delta_{i,A}$. 
\end{itemize}

The minimal setting for comparing how wet-dry and reactant cycles affect chemical processes is a ternary mixture $(i=A, B, S)$ with a single (reversible) uni-molecular chemical reaction between $A$ and $B$. 
For a volume-conserving chemical reaction $A \rightleftharpoons B$, we have $\nu_A=\nu_B \equiv \nu$. 
Moreover, the chemical reaction flux $r_B=-r_A \equiv r$ is governed by the chemical potentials ($\mu_A$ and $\mu_B$) for the two chemical rates.
This reaction flux can be written as~\cite{julicherModelingMolecularMotors1997b}
\begin{equation}
    r(\bm{x},t) = k_c\left[\exp\left\{\frac{\mu_A(\bm{x}, t)}{k_\text{B}T}\right\}-\exp\left\{\frac{\mu_B(\bm{x}, t)}{k_\text{B}T}\right\}\right]\,,    \label{eq:reaction_rate}
\end{equation}
where $k_c$ is the kinetic rate coefficient of the chemical reaction. Chemical equilibrium corresponds to $\mu_A(\bm{x}) = \mu_B(\bm{x})$ at each position $\bm{x}$ in the mixture.

\subsection{Chemical work and power with cycling reservoirs}\label{sect:chem_work}

\begin{figure}[tb]
    \centering  \includegraphics[width=0.5\linewidth]{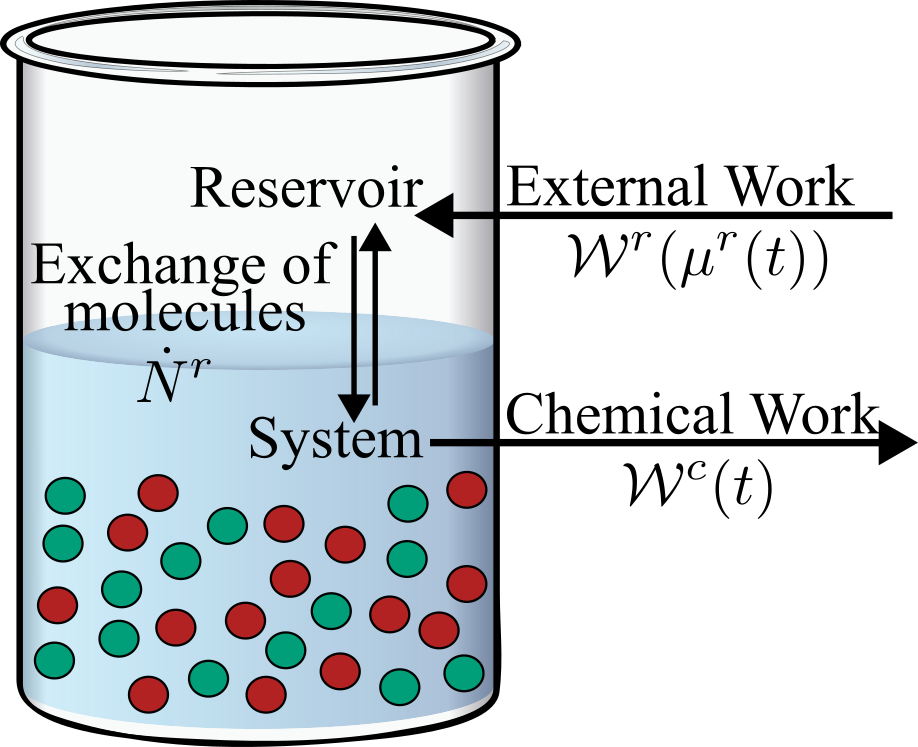}   \caption{\textbf{Illustration of chemical work.} 
    To cycle the chemical potential $\mu^r(t)$ of the reservoir, work $\mathcal{W}^r$ is performed on the reservoir.
    As a result, molecules associated with the chemical potential $\mu_i^r$ are continuously exchanged with the mixture (blue shaded domain), with $\dot{N}^r$ as the exchange rate of such molecules. 
    This exchange of molecules drives the mixture away from equilibrium. Thus, the chemical reaction (here between red and green molecules) is altered, leading to chemical work $\mathcal{W}^c$ performed by the chemical reactions. 
    The harvest of energy or chemical power from the reservoir on the molecular scales is characterized by the efficiency $\eta=\mathcal{W}^c/\mathcal{W}^r$.
} 
    \label{fig:illustration_harvest_work}
\end{figure}

The reservoir chemical potential $\mu^r_i(t)$ is cycled with a frequency $\Omega$ in Eq.~\eqref{rev_osi}, maintaining the system away from chemical equilibrium, resulting in continuous and periodic chemical turnover between the reacting components $A$ and $B$. 
The turnover can be regarded as chemical work performed by the mixture, as depicted in an illustration, Fig.~\ref{fig:illustration_harvest_work}. 
We define the harvested work of the system from the reservoir as the energy dissipated by the chemical reactions over a period $\tau=1/\Omega$ (see Appendix \ref{energy_dissipation_and_work} for details),
\begin{equation}
    \mathcal{W}^c = -\lim_{t_0\rightarrow\infty}\int_{t_0}^{t_0+1/\Omega} \int_{V(t)} \text{d}^3\bm{x} \,\text{d}t \, \sum_i \frac{\mu_i r_i}{\nu_i}\,.
    \label{chem_work}
\end{equation}
For the given reaction scheme, the harvested work from the reservoir takes the form
\begin{equation}
    \mathcal{W}^c = -\lim_{t_0\rightarrow\infty}\int_{t_0}^{t_0+1/\Omega} \int_{V(t)} \text{d}^3\bm{x} \,\text{d}t \, \frac{r 
    \, \Delta \mu_c}{\nu}
    \,,
\label{chem_work2}
\end{equation}
where 
\(\Delta \mu_c = \mu_A - \mu_B\)
is the thermodynamic force of the reaction 
$A \rightleftharpoons B$. No work is performed if the system is at chemical equilibrium (\(\Delta \mu_c= 0\)), or if the reaction flux vanishes (\(r = 0\)).

For the reservoir to undergo cycles, external work must be performed on the reservoir (Fig.~\ref{fig:illustration_harvest_work}):
\begin{equation}
    \mathcal{W}^r = \lim_{t_0\rightarrow\infty} \int_{t_0}^{t_0+1/\Omega} \int_{\partial\Theta_r}\,\text{d}^2\bm{x}\, \text{d}t \, \sum_i \mu_i^r \tilde{\bm{h}}_i\cdot\hat{\bm{n}}. \label{eq:res_work}
\end{equation}
An efficiency $\eta$ can be defined as the ratio of the harvested chemical work per cycle  (Eq.~\eqref{chem_work}) and the work of the reservoir to maintain the cycles (Eq.~\eqref{eq:res_work}):
\begin{equation}\label{eq:sys_eff}
    \eta=\frac{
\mathcal{W}^c}{\mathcal{W}^r}\,.
\end{equation}
In other words, the efficiency $\eta$ characterizes the fraction of external work harvested as chemical work.
Chemical work is expected to increase with longer cycle periods or small frequencies. 
To enable a fair comparison across cycles with different oscillation frequencies, we focus on chemical work per period or average chemical power (see Fig.~\ref{fig:systems_illistration}(c)), defined as 
\begin{equation}
\label{eq:chemical_power}
\mathcal{P}^c = \Omega\, \mathcal{W}^c \, .
\end{equation}
In our work, the harvested average chemical power $\mathcal{P}^c$ and the efficiency $\eta$ are considered as key readouts that characterize the harvesting process.

\subsection{Reduced description at phase equilibrium}\label{eq:theory_phaseeq}

The dynamic Eq.~\eqref{eq:phi_dot} describing the spatial-temporal evolution simplifies under the condition that the diffusive exchange between phases occurs much faster than the other time-scales of the system, i.e., $k_c,k_{r,i},\Omega \ll \Lambda_i k_B T / L^2$. 
Here, $L$ denotes the system size, and the associated reaction-diffusion length scale is given by $\sqrt{\Lambda_i k_B T / k_c}$. 
The spatial gradients become negligible and the moving boundary problem associated with Eq.~\eqref{eq:phi_dot} becomes greatly simplified in this limit. 
Most importantly, the cyclic dynamics can be represented as chemical trajectories in the mixture's thermodynamic phase diagram, as proposed in Ref.~\cite{bauermannChemicalKineticsMass2022d}. 
This representation enables deciphering the mechanism underlying the harvest of chemical power.

When interactions between components are sufficiently strong,  liquid-liquid phase separation can occur.
For two-phase coexistence, and in this limit of fast diffusion, the two phases $(\RN{1},\RN{2})$ remain in local phase equilibrium at all times, such that the chemical potential of each phase balances $\mu_i^\RN{1} = \mu_i^\RN{2}$.
Thus, the composition in each phase is uniform but different, characterized by distinct volume fractions $(\phi_i^\RN{1}, \phi_i^\RN{2})$ and corresponding phase volumes $V^\RN{1}$ and $V^\RN{2}$.
Note that the phase volumes sum up to the total system volume, $V = V^\RN{1} + V^\RN{2}$. 
Thus, the reduced dynamic equations governing chemical reactions in a mixture that is subject to cycling reservoirs are governed by (see Ref.~\cite{bauermannChemicalKineticsMass2022d} for the derivation):
\begin{subequations}
\label{eq:kinetic_evolve}
\begin{align}
\dv{\phi_i^{\RN{1}/\RN{2}}}{t} &= s_i^{\RN{1}/\RN{2}} - j^{\RN{1}/\RN{2}}_i - \phi_i^{\RN{1}/\RN{2}}\frac{1}{V^{\RN{1}/\RN{2}}} \dv{{V}^{\RN{1}/\RN{2}}}{t}, \label{eq:phi_dot_fast_diff}\ \\
\frac{1}{V^{\RN{1}/\RN{2}}} \dv{{V}^{\RN{1}/\RN{2}}}{t} &= \sum_{k=A,B,S} \left( s_k^{\RN{1}/\RN{2}} - j_k^{\RN{1}/\RN{2}} \right), \label{eq:vol_change}
\end{align}
\end{subequations}
where $i = A, B$, and the diffusive fluxes $j_i^{\RN{1}/\RN{2}}$ maintaining the system as  phase coexistence.

The combined change from reactions and boundary fluxes in Eq.~\eqref{eq:phi_dot} combines to a single source term in Eq.~\eqref{eq:kinetic_evolve}:
\begin{equation}\label{eq:si_homo}
    s_i^\alpha = r_i^\alpha + {h}^\alpha_i \,.  
\end{equation}
and describe the net source-flux in each  spatially homogeneous phase $\alpha\in\{\RN{1},\,\RN{2}\}$. 
The volume fraction reservoir flux is related to the surface area flux density given in Eq.~\eqref{eq:boundary_flux} by $h_i^\alpha = \mathcal{A} \nu_i \hat{\bm{n}} \cdot \tilde{\bm{h}}_i^\alpha / V^\alpha $, finding
\begin{equation}
\label{eq:boundary_flux_integrated}
    h_i^\alpha = \frac{\nu_i k_{r,i}^\alpha}{V^\alpha}\left[\exp\left\{\frac{\mu_i^r}{k_\text{B}T}\right\}-\exp\left\{\frac{\mu_i}{k_\text{B}T}\right\}\right] \, . 
\end{equation}
We defined the reservoir exchange rate coefficient $k_{r,i} = \mathcal{A} \, \tilde{k}_{r,i}$ with units [1/time], where $\mathcal{A}=|\partial \Theta_r|$ is the total surface area connected to the reservoir. 
Note that the chemical potential has no superscript indicating the phase I or II since both phases are at phase equilibrium with
$\mu_i \equiv\,\mu_i^\RN{1}\, = \mu_i^\RN{2}$.

The reaction flux in each phase reads
\begin{equation}\label{eq:reaction_rate2}
    r^\alpha = k_c^\alpha\left[\exp\left\{\frac{\mu_A}{k_\text{B}T}\right\}-\exp\left\{\frac{\mu_B}{k_\text{B}T}\right\}\right]\,,
\end{equation}
such that $r_B^\alpha=-r_A^\alpha\equiv r^\alpha$, and 
chemical equilibrium corresponding to $\Delta \mu_c = \mu_A - \mu_B =0$. 

When the system can phase separate, both the reaction and reservoir exchange rate coefficients can, in general, be different between the phases. For simplicity, we consider throughout this work that the reaction rate coefficient is not phase-dependent, $k_c^\RN{1} = k_c^\RN{2}=: k_c$. 
Moreover, either phase I or phase II is connected to the reservoir through the surface area  $\mathcal{A}$. 
Say the phase I is connected to the reservoir, $k_r^\RN{1} =: k_r$, while consistently, $k_r^\RN{2} =0$. 
We note that, due to phase equilibrium, the choice of which phase is in contact with the reservoir is conceptually irrelevant. 
However, in experiments, the reservoir exchange rate coefficients $k_r$ were shown to differ significantly between the phases~\cite{aranberri2002emulsions, cerretani2013water}.

Spatial averages, as used in the definition of the chemical work in Eq.~\eqref{chem_work}, reduce to simple phase averages for the kinetics at phase equilibrium since spatial profiles are homogeneous in each phase. 
For a general quantity $X^\alpha(t)$, the phase average is given as
\begin{equation}
    \bar{X}(t) = \frac{V^\RN{1}(t)X^\RN{1}(t) + V^\RN{2}(t)X^\RN{2}(t)}{V(t)}\,.
\end{equation}
Note that the volume of each phase $V^\RN{1}$ and $V^\RN{2}$ is dynamic and changes according to Eq.~\eqref{eq:vol_change}. 
As a result, the total system volume $V=V^\RN{1}+V^\RN{2}$ changes in time as well. 
For example, the system volume $V$ can decrease during the drying period as solvent molecules evaporate. 
In general, the exchange with the reservoir can dilute or concentrate the system's composition by adding/removing solvent or reactant molecules.
This process is illustrated in Fig.~\ref{fig:systems_illistration}(a).

\section{Results}
\begin{figure*}[tb]
    \centering
    \makebox[\textwidth]{\includegraphics[width=\textwidth]{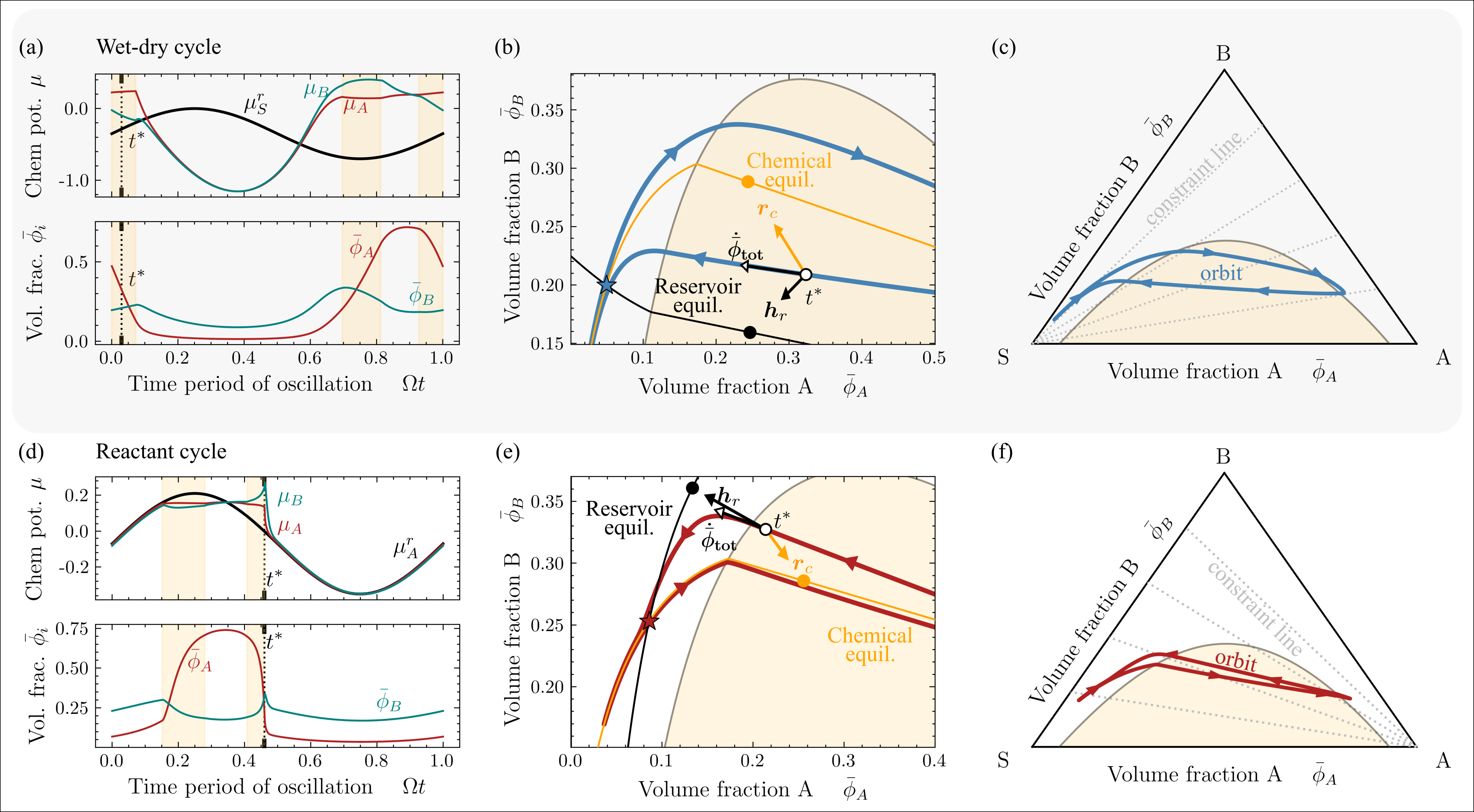}}    \caption{\textbf{Reactant and wet-dry cycles can induce oscillations of the mixture's composition  with intermittent periods of phase separation.}
    (a,d)  Periodically exchanging molecular components with the reservoir can lead to intermittent periods of phase separation, highlighted by yellow domains. 
    While the chemical potential of the mixture is maintained close to the oscillating value of the reservoir (top), 
    the volume fractions of the components $A$ and $B$, exhibit oscillations containing multiple frequencies.
    (b,e) At any given moment along the trajectory, the mixture's composition  experiences two effective driving fluxes: one from the external reservoir \(\boldsymbol{h}_r\,\equiv\,(\bar h_A,\bar h_B)\) (black arrow) and another from chemical reactions \(\boldsymbol{r}_c\,\equiv\,(r_A,r_B)
    = (-r,r)\) (orange arrow). Each of the force components points towards its respective equilibrium composition, which is shown by the black dot (reservoir equilibrium) or orange dot (chemical equilibrium), respectively. As a result,  the system evolves in a net direction, i.e, \(\boldsymbol{\dot\phi}_\text{tot}=\bm{h}_r\,+\,\bm{r}_c\) (Eq.~\eqref{eq:phitotdot}).
    (c,f) After multiple oscillations, the system's trajectory forgets its initial condition and settles into a  stable and closed loop in the thermodynamic phase diagram, shortly referred to as ``orbit''.
    }
    \label{fig:orbit_shape_and_flux}
\end{figure*}

\subsection{Phase diagram orbits generated by cycling reservoir}\label{Reservoir_cycles_lead_to_phase_diagram_orbits}

\subsubsection{Chasing thermodynamic equilibrium}\label{chasing_thermodynamic_equilibrium}

The cycling reservoir leads to closed-loop trajectories (orbits) in the thermodynamic phase diagram, as shown in Fig.~\ref{fig:orbit_shape_and_flux}(c,f).
A fundamental point determining the position of the orbit in the phase diagram is thermodynamic equilibrium. 
Thermodynamic equilibrium in the considered ternary mixture corresponds to a unique composition of average volume fractions $(\bar{\phi}_A, \bar{\phi}_B)$, marked by a star in the phase diagrams shown in Fig.~\ref{fig:orbit_shape_and_flux}(b,e).
Graphically, this thermodynamic equilibrium point is the intersection of the reservoir equilibrium (black) line $\mu_i^r(t) = \mu_i$ and chemical equilibrium (orange) line  $\mu_A = \mu_B$. 
The mean thermodynamic equilibrium is defined by $\mu_i^r(t) = \left<\mu_i^r\right>$, and represents a relevant stationary reference case to compare to the system connected to the cycling reservoir. In fact, the orbit always encloses the mean thermodynamic equilibrium point.
Due to the cycling reservoir $\mu_i^r(t)$, the thermodynamic equilibrium point is not stationary and undergoes periodic motion in the phase diagram. 
And since the reacting mixture aims to relax to thermodynamic equilibrium, the average volume fraction $(\bar{\phi}_A, \bar{\phi}_B)$ chases the non-stationary thermodynamic equilibrium point. \par

The shape of the orbits is set by special attractors in the phase diagram. 
At an arbitrary time \(t=t^*\), the system's phase average volume fraction \(\{\bar{\phi}_i\}\)  (marked by a white dot, see Fig.~\ref{fig:orbit_shape_and_flux}(b,e)) changes due to fluxes toward two distinct attractors in phase space:
One attractor is set by the reservoir (black dot, \(\mu_i = \mu_i^r\)), and the other attractor is set by the chemical reaction (orange dot, \(\mu_A = \mu_B\)). The net flux $\text{d}\bm{\bar\phi}_\text{tot}/\text{d}t$ from the respective attractors are given as,
\begin{equation}
\label{eq:phitotdot}
\text{d}\bm{{\bar\phi}}_\text{tot}/\text{d}t=\,\bm{r}_c + \bm{h}_r \,,
\end{equation}
where $\bm{{\bar\phi}}_\text{tot}=(\bm{\bar\phi}_A, \bm{\bar\phi}_B)$ is the state vector of the volume fractions.
The fluxes $\bm{r}_c$ and $\bm{h}_r$ point to their respective attractors as shown in Fig.~\ref{fig:orbit_shape_and_flux}(b,e), which are set by their specific constraints.
Specifically, the chemical reaction flux $\bm{r}_c\equiv(-r,\,r)$ (orange arrow) conserves the total volume fraction $\bar{\psi}=\bar{\phi}_A +\bar{\phi}_B$ of the reacting species, while the reservoir fluxes $\bm{h}_r\equiv (\bar h_A, \bar h_B)$ (black arrow) conserves the average volume fraction ratio \(\bar{\varphi}_i\) of all the other components as \(\bar{\varphi}_i = \bar{\phi}_{j\neq i}/\bar{\phi}_{k\neq i}\). 
The constraint lines for the two different reservoirs are shown in dotted grey in Fig.~\ref{fig:orbit_shape_and_flux}(c,f). 
For wet-dry cycles, the ratio \(\bar{\varphi}_S = \bar{\phi}_B/\bar{\phi}_A\), and for reactant ($A$) cycle, \(\bar{\varphi}_A = \bar{\phi}_S/\bar{\phi}_B\) remain unaffected by external reservoir. 
The system's composition in phase space moves in the direction of the net flux $\text{d}\bm{\bar\phi}_\text{tot}/\text{d}t$, indicated by a white arrow (tangential to the orbit), which is unconstrained. 
The reservoir chemical potential $\mu_i^r(t)$ changes in time given by Eq.~\eqref{rev_osi}, and the thermodynamic equilibrium changes concomitantly. 
This makes the thermodynamic equilibrium oscillate along the chemical equilibrium line; see movies in the Supplementary Material~\cite{SI_ref}.
The system, in turn, chases this dynamic equilibrium, with the reservoir repeatedly pulling it away from chemical equilibrium.
During its trajectory, the system minimizes its free energy by phase separating into phases I and II, which differ in their compositions. 
The binodals encompass these phase compositions and  are shown in yellow in 
Fig.~\ref{fig:orbit_shape_and_flux}. 
After multiple reservoir cycles, the system's phase diagram trajectories become periodic, forming closed loops referred to as orbits in the following. 

\subsubsection{Phase diagram orbits are set by cycle frequency}\label{Phase_diagram_orbits_are_set_by_cycle_frequency}

Starting from an arbitrary initial composition $(\bar{\phi}_A(t=0), \bar{\phi}_B(t=0))$, the mixtures trajectories after multiple reservoir cycles converge to a periodic orbit in the phase diagram with a shape distinct to the system's parameters, such as the cycle frequency $\Omega$ (Eq.~\eqref{rev_osi}). 
For frequencies \(\Omega/k_c \ll 1\), the changes in the reservoir chemical potential are slow compared to the chemical reactions in the system. 
Thus, the system has enough time to relax toward the thermodynamic equilibrium point. During this rapid relaxation, the system experiences only a little exchange of molecules with the external reservoir, and the orbits are very close to the chemical equilibrium line for both reactant and wet-dry cycles. 
Meanwhile, for frequencies \(\Omega/k_c \gg 1\), the reservoir cycles are much faster than chemical reactions. 
Thus, the system is too slow to respond to the changes made by the reservoir, making the system's volume fraction majorly evolve along a respective reservoir constraint line (\(\bar{\varphi}_S\) for wet-dry cycle, and \(\bar{\varphi}_A\) for reactant cycle). 
The orbits take an elliptical shape with the major axis aligned along their specific constraint line. 
Between these two extremes, a frequency exists at which the system has sufficient time to react to the external reservoir and can be driven away from the chemical equilibrium line, thereby maximizing the chemical power of a cycle (Eq.~\eqref{chem_work} and Eq.~\eqref{eq:chemical_power}).

\subsection{Maximal power is enhanced by increasing the amplitude of the external reservoir}\label{tuning_power_amplitude}

In this section, we analytically determine the dependence of the maximal power on the amplitude of the reservoir oscillations.
For simplicity, we focus on a homogeneous system.
For the analytics, we consider small amplitudes of the cycling reservoir (\(\hat \mu/k_\text{B}T\,\ll\,1\)). 
In this case, we will demonstrate that the orbits are close to the mean thermodynamic equilibrium, allowing for a linearization of the system's dynamics around this point.
We will show that the system can then be mapped onto a driven harmonic spring network.

\subsubsection{Lower reservoir amplitude narrows orbits toward mean thermodynamic equilibrium}\label{Lower_reservoir_amplitude_narrows_orbits_toward_mean_thermodynamic_equilibrium}

\begin{figure*}[tb]
    \centering
    \makebox[\textwidth]{\includegraphics[width=\textwidth]{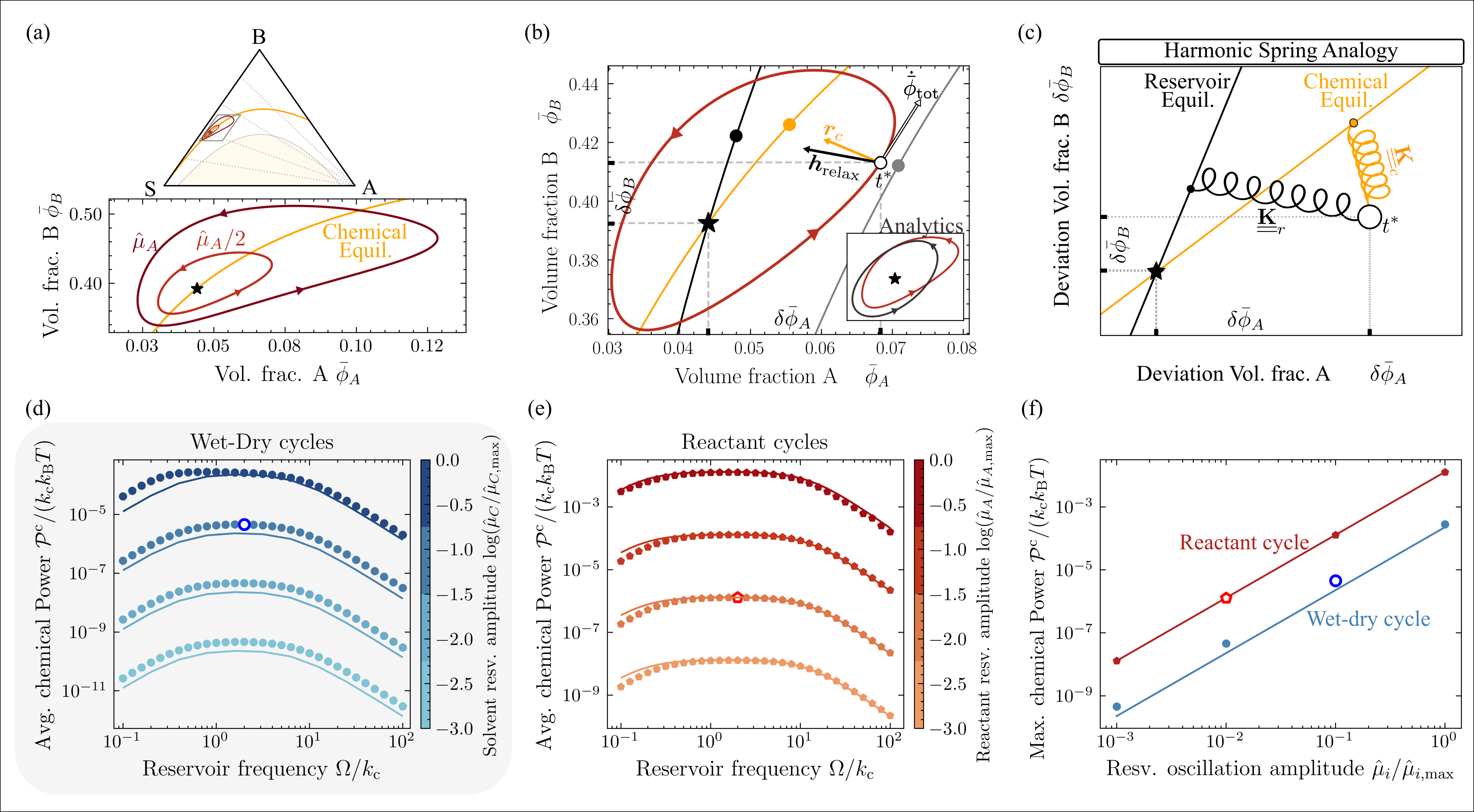}}
     \caption{\textbf{
     The oscillating chemical and reservoir equilibria 
     act like compositional harmonic springs in the thermodynamic phase diagrams.}  
     (a) Decreasing the amplitude of reservoir cycles $\hat{\mu}_A$, decreases the orbit size. 
     (b) Each point along the orbit (red line), system (open white symbol) experiences a net flux \(\bm{\dot{\phi}}_\text{tot}\) which is tangent to the orbit. \(\bm{r}_c\) is the flux due to chemical reaction. \(\bm{h}_r\) is the flux from the reservoir mean chemical potential. 
     Inset: Comparison between orbit obtained from numerical solution (red) and analytic calculation using the mapping on a two-spring model (black) agrees well with each other.
     (c) Same system and time point as in (b) but illustrated using a spring model with elongations
     $\delta \bar\phi_A$ and $\delta \bar\phi_B$. 
     The harmonic springs are characterized by the spring matrices $\mat{\mathbf{K}}_c$ and $\mat{\mathbf{K}}_r$ related to chemical reactions and the reservoir. They exert effective forces directed toward the reservoir and the chemical equilibrium point at each. 
     (d,e) The larger the amplitude of the respective oscillating reservoir $\hat{\mu}_A$, the larger the average chemical power $\mathcal{P}^c/(k_c k_\text{B}T)$, while the resonance frequency of the maximal power remains approximately unchanged. The data points are results from numerical simulations, while the solid lines represent analytic predictions using the two-spring model. 
     (f) The maximum chemical power increases linearly with increasing oscillation amplitude.}
    \label{fig:amplitude_change}
\end{figure*}

Decreasing the cycling amplitude $\hat\mu_i/k_\text{B}T$ decreases the orbit size. 
The system's average volume fraction deviation from the mean thermodynamic point (black star) in the phase diagram becomes smaller as shown in the compositional phase diagram in Fig.~\ref{fig:amplitude_change}(a).  
For a low amplitude cycle (\(\hat \mu_i/k_\text{B}T\,\ll\,1\)), we can decouple the flux from the reservoir \(\bm{h}_r\) into relaxation flux \(\bm{h}_\text{relax}\) and the driving flux $\bm{h}_\text{driving}$
(full expression see 
Eq.\eqref{eq:fluxes_in_relax_driving} and derivations in shown on Appendix~\ref{analytical_orbits_for_small_amplitude}):  
\begin{equation}
\bm{h}_r = \bm{h}_\text{relax}+\,\bm{h}_\text{driving} \, . 
\end{equation}
The reservoir flux \(\bm{h}_\text{relax}\) corresponds to the flux without reservoir cycles  ($\bm{h}_r(\hat\mu_i =\,0)$). 
It pulls the system's trajectory towards its mean reservoir equilibrium with $\mu_i^r = \left<\mu_i^r\right>$. 
Moreover, \(\bm{h}_\text{driving}\) is the flux due to the driving of the cyclic reservoir chemical potential, pointing towards the attractor point (grey symbol) on the reservoir equilibrium line (grey line). 
Both \(\bm{h}_\text{relax}\) and \(\bm{h}_\text{driving}\) are parallel along their constraint line \(\bar\varphi_i\) of the reservoir.
As the reservoir is cycled, \(\bm{h}_\text{driving}\) maintains the system away from equilibrium.
As the cycle amplitude \(\hat\mu_i\) becomes smaller, the reservoir equilibrium (grey line) oscillates less from the mean equilibrium (black line). 
As a result, we see that the system deviates less from mean thermodynamic equilibrium, leading to orbits that are closer to the mean thermodynamic equilibrium point.
By cycling the reservoir by small amplitude ($\hat{\mu}_i/k_\text{B}T\ll1$) and choosing the mean thermodynamic equilibrium outside the binodal, we can make sure that the orbit never enters the binodal, thereby fulfilling the assumption for the subsequent analytics of a homogeneous system at all times. 

\begin{figure*}[tb]
    \centering
    \makebox[\textwidth]{\includegraphics[width=\textwidth]{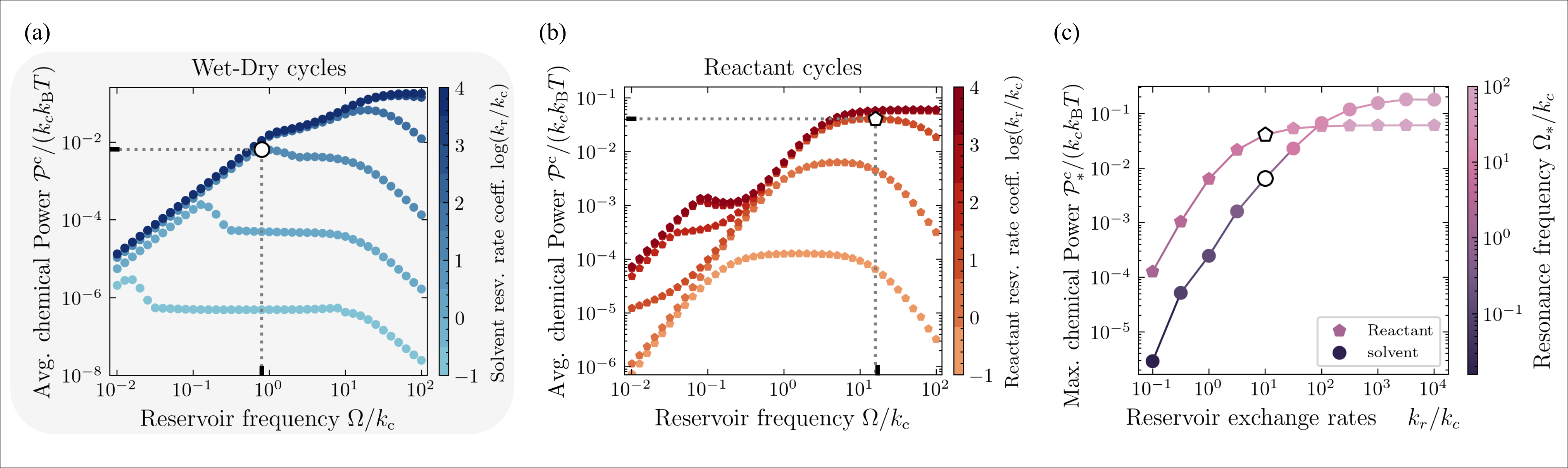}}
 \caption{\textbf{Larger exchange rates $k_r$ between reservoir and mixture enhance harvested chemical power.} 
 (a,b) There exists a resonance frequency that maximizes the chemical power. This resonance frequency is strongly affected by changing the reservoir kinetic rates. 
 (c) Maximal chemical power increases and saturates for both types of reservoir cycles. 
 Interestingly, there is a crossover below which reactant cycles lead to more chemical power, while above wet-dry cycles dominate.}
 \label{fig:kinetic_rate_variation}
\end{figure*}

\subsubsection{System behaves as a driven spring network}
\label{System_behaves_as_a_driven_spring_network} 
The dynamics of the reacting mixture, subject to a reservoir that can exchange particles  with the mixture, correspond to two coupled and driven oscillators elongating and retracting in the mixture's thermodynamic phase diagram. 
For cycles with small reservoir amplitudes (\(\hat\mu_i\ll k_\text{B}T\)), the system's orbital trajectories get closer towards the mean thermodynamic equilibrium $\{\bar\phi_i^*\}$ (black star, Fig.~\ref{fig:amplitude_change}(a)).
As the deviations of systems' average volume fractions from the mean thermodynamic equilibrium (i.e, $\delta\bar\phi_i = \bar\phi_i - \bar\phi_i^*$) decrease with smaller cycling amplitude $\hat{\mu}_i$,  we expand the fluxes in average volume fractions around the mean thermodynamic point; see Appendix~\ref{analytical_orbits_for_small_amplitude} for more details on the linearization.
The linearized time evolution of average volume fraction deviations $\delta\bar\phi_i$ from the mean equilibrium point then takes the form of a driven harmonic spring network:
\begin{equation}
\label{linear:ode's}
    \dv{\delta \bm{\Phi}(t)}{t}
    = -\mat{\mathbf{\Gamma}}\cdot \delta \bm{\Phi}(t) + \mathbf{R}(t) \,  , \text{with} \, \, \delta\bm{\Phi}(t) = 
    \begin{pmatrix}
    \delta\bar{\phi}_A\\[0.5em]
   \delta\bar{\phi}_B
\end{pmatrix} \, . 
\end{equation}
The second term $\mathbf{R}(t)$ (defined in Eq.~\eqref{eq:driving_vec} and discussed in Appendix~\ref{analytical_orbits_for_small_amplitude}) is a driving term stemming from the cyclic reservoir that pulls the system away from equilibrium. 
The first term in the above Eq.~\eqref{linear:ode's} pulls the system towards its mean thermodynamic equilibrium with the damping matrix $\mat{\mathbf{\Gamma}} = \mat{\mathbf{M}}\cdot\mat{\mathbf{K}}$.
Here, $\mat{\mathbf{M}}$ is an effective mobility and $\mat{\mathbf{K}}$ is an effective spring matrix that are defined by the linear response relaxation kinetics (see Appendix~\ref{app:mobilities_and_forces}). 
The relaxation can be understood as a coupled harmonic spring network where each spring is attached to a dynamic point along the respective equilibrium lines; see Fig.~\ref{fig:amplitude_change}(c).
As a result, the system (white symbol) is subject to spring forces in the thermodynamic phase diagram (white symbol) with springs characterized by spring matrices $\mat{\mathbf{K}}_c$ for chemical reactions and $\mat{\mathbf{K}}_{r,x}$ for the precursor ($x \equiv A$) or solvent ($x = S$) reservoir. 
The farther the system is displaced from its rest length (in the phase diagram), the larger is the resultant force:
\begin{equation}
\label{eq:force_chem_resv}
    \mathbf{F}_c = -\mat{\mathbf{K}}_c\cdot\delta\bm{\Phi}\,,\quad \mathbf{F}_{r,x} = -\mat{\mathbf{K}}_{r,x}\cdot\delta\bm{\Phi}\,,
\end{equation}
where $\mathbf{F}_c$ and $\mathbf{F}_{r,x}$ are the forces due to chemical and reservoir springs, respectively.
The corresponding mobility matrices are defined as $\mat{\mathbf{M}}_c$ and $\mat{\mathbf{M}}^x_{r}$.
These coupled spring matrices together with their mobilities, enter the effective damping coefficient as,
\begin{align}
    \mat{\mathbf{\Gamma}} &= \mat{\mathbf{\Gamma}}_c + \mat{\mathbf{\Gamma}}_{r,x}\nonumber\\[0.5em]
    &=\mat{\mathbf{M}}_c\cdot\mat{\mathbf{K}}_c + \mat{\mathbf{M}}^x_r\cdot\mat{\mathbf{K}}_{r,x}\,.
\end{align}
The expressions for both the chemical and reservoir spring matrices are given in Appendix~\ref{app:mobilities_and_forces}, specifically Eq.~\eqref{eq:mat_chem_spring_const} and Eq.~\eqref{eq:mat_resv_spring_const}.

The deviation from  system volume at the equilibrium, $\delta V$, changes as follows
\begin{equation}
\label{eq:vol_change_around_mean}
    \delta\dot V = V^*(h_{\text{relax},x} + h_{\text{driving},x})\,.
\end{equation}
Here, $h_{\text{driving},x}(t)$ is a function of time due to the external cyclic reservoir, and $h_{\text{relax},x}(\delta\bm{\Phi}(t))$ depends on the deviations in volume fractions at a time $t$ (see Appendix~\ref{analytical_orbits_for_small_amplitude} for derivation and Eq.~\eqref{eq:linear_volume_flux_2} for the full expression).
We note that the fluxes in Eq.~\eqref{eq:vol_change_around_mean} do not depend on $\delta V$
to leading (linear) order term.
To determine the change in system volume $\delta V(t)$ around the system volume at the mean equilibrium $V^*$, we integrate Eq.~\eqref{eq:vol_change_around_mean} in time over a period $t\in\,[0, 1/\Omega]$.\par

We found out that the deviation from the mean thermodynamic point $\delta \bm{\Phi}(t)$  is a function of cycling frequency $\Omega$ (see Appendix~\ref{analytical_orbits_for_small_amplitude}, see (Eq.~\eqref{analytics:sol_4}) for the functional form).
The analytical solution for an exemplary case is shown in Fig.~\ref{fig:amplitude_change}(b) as an inset (black line), indicating a good agreement with the numerical solutions of the full non-linear system  (Eq.~\eqref{eq:kinetic_evolve}; red line in inset).
From the analytic solutions, we calculated the average chemical power $\mathcal{P}^c$
(see Appendix~\ref{analytical_power_output} and (Eq.~\eqref{eq:analytical_power}) for the full analytic expression for chemical power). 
We find that the average chemical power extracted from the wet-dry and reactant cycles, shown as a function of frequency \(\Omega\), agrees well with the numerical solutions (Fig.~\ref{fig:amplitude_change}(d,e)).

\subsubsection{System produces maximum power at resonance frequency } \label{selection}

There exists a reservoir frequency for which the average chemical power is maximal. For slow cycles (\(\Omega \ll k_c\)), the systems' average volume fraction is close to chemical equilibrium, \(\Delta \mu_c \simeq 0\). In contrast, for fast cycles (\(\Omega \gg k_c\)), the system response is much slower than the changes of the reservoir, resulting in a small chemical flux \(r_c\); see Eq.~\eqref{eq:reaction_rate2}. Thus, the system's chemical power \(\mathcal{P}^c\) becomes negligible in both frequency limits. However, for intermediate frequencies, the system has enough time to respond to the reservoir, pulling it away from chemical equilibrium, harvesting more chemical power $\mathcal{P}^c$.
Strikingly, this creates a distinct frequency \(\Omega_*\) at which the chemical power \(\mathcal{P}^c_*\) is maximal. This frequency corresponds to a resonance of the bulk with the reservoir, and is referred to as a resonance frequency \(\Omega_*\), indicated as a white point in Figs.~\ref{fig:amplitude_change}(d,e).

Wet-dry and reactant cycles in general show different resonance frequencies \(\Omega_*\) and different maximal power at resonance \(\mathcal{P}^c_*\). 
Interestingly, the resonance frequency \(\Omega_*\) is independent of the amplitude \(\hat{\mu}_i\) of the oscillating reservoir; see Fig.~\ref{fig:amplitude_change}(d,e). 
Moreover, both cycling reservoirs give rise to a chemical power with a quadratic increase at low frequencies \(\mathcal{P}^c\propto \Omega^2\).
In contrast,  \(\mathcal{P}^c\propto \Omega^{-2}\) in the limit of large frequencies. 
The average chemical power in the linearized regime is derived in {Appendix~\ref{analytical_power_output}} and given in Eq.~\eqref{eq:analytical_power}. 
This analytical expression is plotted as solid lines in Fig.~\ref{fig:amplitude_change}(d, e).
We see that the analytics captures well both scaling behaviors at low and high frequencies.

\begin{figure*}[tb]
    \centering
    \makebox[\textwidth]{\includegraphics[width=\textwidth]{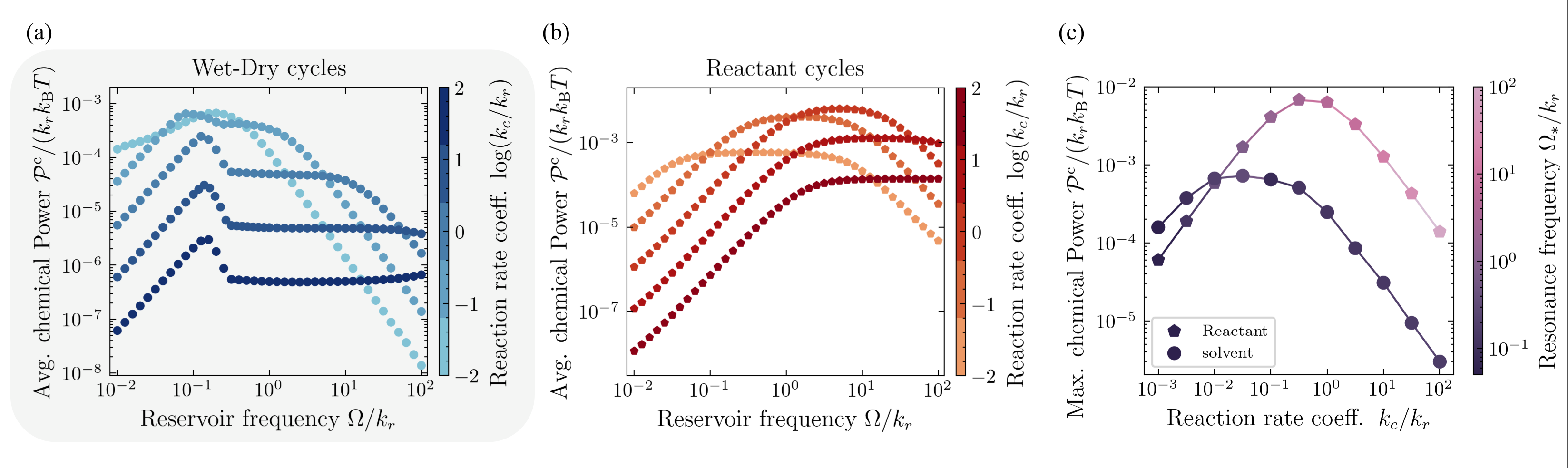}}
 \caption{\textbf{Slow reactions can increase the maximum chemical power.}
    (a,b) For different reaction rates $k_c$ (colorbar), the average chemical power is shown upon varying cycling frequency. 
    There exists a resonance frequency $\Omega_*/k_r$ that maximizes the average chemical power.
    (c) The maximum chemical power $\mathcal{P}^c_*/(k_ck_\text{B}T)$ increases and then decreases upon increasing reaction rate $k_c/k_r$.
    There exists an optimal reaction rate that maximizes the maximal power for both Wet-dry and reactant cycles.}
\label{fig:exchange_rate_variation}
\end{figure*}

\subsubsection{Maximal chemical power scales with cycle amplitude}
\label{Maximal_work_scales_with_cycle_amplitude}

More chemical power can be harvested by increasing the oscillation amplitude of the reservoir. The maximum chemical power \(\mathcal{P}^c_*\), i.e, the chemical power at resonance frequency \(\Omega_*\), is depicted for different amplitudes \(\hat\mu_i\) in Fig.~\ref{fig:amplitude_change}(f). For a fair comparison between wet-dry and reactant cycles, both share a common mean thermodynamic equilibrium point with the same oscillation amplitude.  
As shown in Fig.~\ref{fig:amplitude_change}(f), the harvested chemical power increases linearly with increasing reservoir amplitude \(\hat\mu_i\). Here, for the parameter considered, reactant cycles lead to larger chemical work than the wet-dry cycle, approximately by a factor of 10. However, this is not generally true, and it depends on the interactions of the system, as will be seen in the following section.

\subsection{Maximum power is enhanced by increasing the kinetic rates of the external reservoir}\label{tuning_power_kinetic_rate}

Resonance of solvent reservoirs occurs at much slower frequencies than for reactant reservoirs.
In general, a reservoir can drive the system's average composition into the binodal, as depicted in Fig.~\ref{fig:orbit_shape_and_flux}(b,e). Varying the cycle frequency yields different orbits, resulting in different average chemical power. 
Average chemical power $\mathcal{P}^c/(k_ck_\text{B}T)$ is plotted as a function of cycling frequency \(\Omega/k_c\) for a reactant-phobic interaction, i.e, reactant $A$ phase separates from solvent; see Fig.~\ref{fig:kinetic_rate_variation}(a,b). 
In the previous section, where system composition was homogeneous for all orbits, we observed that the chemical power exhibits a unique resonance frequency that maximizes power. Strikingly, phase separation can give rise to multiple local maxima of chemical power for both reactant and wet-dry cycles, see Fig.~\ref{fig:kinetic_rate_variation}(a,b). For the wet-dry cycle, we show the local (light blue)/global (dark blue) resonance frequencies; see Fig.~\ref{fig:lo_gl_max} in Appendix~\ref{local_global} for different reservoir kinetic rates \(k_r/k_c\). 
For all kinetic rates, the chemical power decays in both low- and high-cycling frequency regimes. 
This behavior arises due to the reasons explained in the previous section. 
The average chemical power saturates for large values of kinetic rates, as shown in Figs.~\ref{fig:kinetic_rate_variation}(a,b).

To harvest more chemical power by cycling the reservoir, we find that it is better to perform reactant cycles for lower reservoir kinetic rates \(k_r\) (or, equivalently, for faster reactions) and better to perform solvent cycles for higher reservoir kinetic rates (or slower chemical reactions). 
This result is evidenced in Fig.~\ref{fig:kinetic_rate_variation}(c), which shows the maximum chemical power \(\mathcal{P}^c_*/(k_ck_\text{B}T)\) as a function of the reservoir kinetic rates \(k_r/k_c\) for both cycles. 
We observe a switch frequency above which wet-dry cycles give a larger maximal chemical power than the reactant cycles. 

\subsection{Harvested chemical power is maximal a distinct reaction rate}\label{Harvested_chemical_power_can_be_increased_by optimal_reaction_speed}

\begin{figure}[tb]
    \centering
    \makebox[\textwidth]{\includegraphics[width=0.85\textwidth]{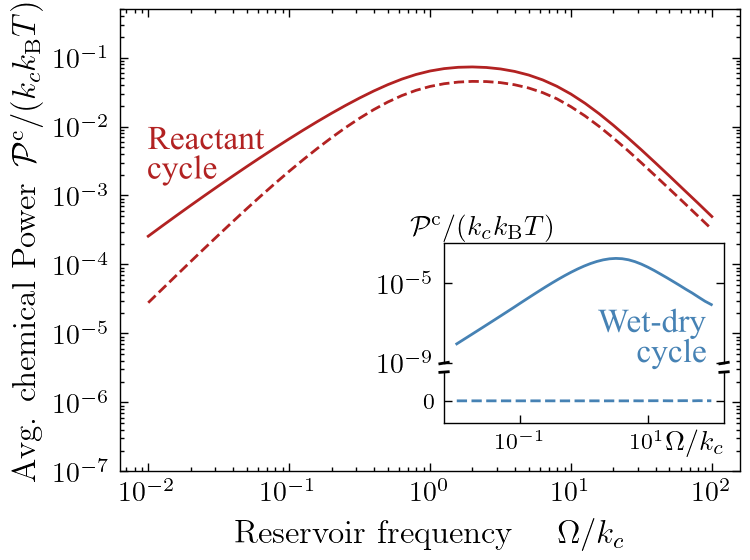}}
   \caption{\textbf{Interactions between mixture's components can lead to enhancement of chemical power:} Chemical power upon varying cycling frequency is shown for both mixtures: (i) with weak (solid lines) and (ii) without interactions (dashed lines). Chemical power can increase due to interactions between the mixture. Inset: Wet-dry cycles acting on an ideal mixture (with no interactions) result in no chemical power. In contrast, the reactant cycle produces chemical power in the case of an ideal mixture.}        \label{fig:dilute_system}
\end{figure}

When varying the 
cycling frequency $\Omega$, we find that there exists an optimal cycling frequency $\Omega_*$ 
at which the chemical power is maximal with the value  $\mathcal{P}^c_*/(k_rk_\text{B}T)$. For a given reaction rate $k_c/k_r$, the average chemical power $\mathcal{P}^c/(k_rk_\text{B}T)$ shows a similar qualitative behavior when we vary the cycling frequency $\Omega/k_r$ as we had seen in our studies earlier. The chemical power decreases for slow and high frequencies. In between there exists a resonance frequency $\Omega_*/k_r$ for which the harvested chemical power takes the maximal value  $\mathcal{P}^c_*/(k_rk_\text{B}T)$; see Fig.~\ref{fig:exchange_rate_variation}(a,b). The maximum chemical power can be tuned by the reaction rate $k_c$.

The maximal harvest of chemical power is largest at a distinct value of the reaction rate $k_c$; see Fig.~\ref{fig:exchange_rate_variation}(c). 
This behavior occurs for both wet-dry and reactant cycles.
For lower and higher reaction rates, the maximal harvested power decreases. 
In the limit of very slow chemical reactions ($k_c/k_r\,\ll\,1$), the harvested power vanishes to zero. 
Using our analytics, we have found that in the limit of slow reactions, the maximal chemical power $\mathcal{P}^c(\Omega)\propto k_c$ scales linearly with the reaction rate $k_c$, see Eq.\eqref{eq:power_lin_2} in Appendix~\ref{analytical_power_output}. 
For very fast reactions ($k_c/k_r\,\gg\,1$), the harvested power vanishes as well. 
This vanishing power originates from reactions being much faster than the exchange of molecules with the reservoir.
In this case, the system is almost at chemical equilibrium ($\Delta\mu_c = 0$) at all times, leading to zero chemical work in power (Eq.~\eqref{chem_work2}). 
Thus, an optimal reaction rate exists that can enhance the maximum power harvested. 
The resonance frequency $\Omega_*/k_r$ is unaffected for the wet-dry cycle type; however, it varies strongly for reactant cycles; see colorbar in Fig.~\ref{fig:exchange_rate_variation}(c). 
Interestingly, we see that the wet-dry cycle can outperform more chemical power than reactant cycles in the limits when reactions are slower than reservoir exchange $k_c\ll k_r$ and the reactant cycles can extract more chemical work from a system with reactions much faster than the exchange of reservoirs, $k_c\gg k_r$.

\subsection{Harvesting chemical power through
wet-dry cycles require non-dilute mixtures}

Chemical power can always be harvested from mixtures undergoing a chemical reaction $A \rightleftharpoons B$ by performing reactant cycles, also when the reactants are dilute. 
Dilute means that interactions among the components (here: $A$, $B$, and solvent) are negligible. 
For reactant cycles, the average chemical power $\mathcal{P}^c(\Omega)$ as a function of reservoir frequency $\Omega$ is quantitatively different from its non-dilute counterpart, but still has a qualitatively similar shape. Moreover, values of power and frequency are in the same order of magnitude for dilute and non-dilute mixtures (Fig.~\ref{fig:dilute_system}).\par

However, the harvest of chemical power through
wet-dry vanishes when the mixture is dilute (Fig.~\ref{fig:dilute_system}, inset). 
This interesting behavior can be understood as follows.
The chemical potential of each component \(i\) in a dilute mixture is \(\mu_i = \mu_i^\circ\ + k_\text{B}T\log(\bar\phi_i)\), where \(\mu_i^\circ\) is some reference chemical potential. The condition of chemical equilibrium is, \(\Delta \mu_c = \mu_A - \mu_B = 0, \) corresponding to $\bar\phi_B/\bar\phi_A =\,\exp{\{(\mu_A^\circ - \mu_B^\circ)/k_\text{B}T\}}$. 
The constraint line for wet-dry cycles is given by $\bar\varphi_S = \bar\phi_B/\bar\phi_A$.
On long time-scales, the chemical equilibrium selects a specific constraint line defined by $\bar\varphi_S=\exp{\{(\mu_A^\circ - \mu_B^\circ)/k_\text{B}T\}}$. 
Thus, on long time-scales, the chemical equilibrium line for the dilute system is parallel to the constraint line of the solvent reservoir and the direction of the reservoir flux, $\bm{h}_r$ (Eq.~\eqref{eq:phitotdot}), points along this constraint line. Therefore, the solvent reservoir oscillates the system along the chemical equilibrium line, causing it to move back and forth. 
In other words, the lack of interactions among the components prevents maintaining the system away from chemical equilibrium, resulting in zero chemical power for all cycling frequencies \(\Omega/k_c\).  
In contrast, for the reactant cycle, we observe that the constraint line of the reservoir, \(\bar\varphi_A = \bar\phi_S/\bar\phi_B\), is not parallel to the chemical equilibrium line, even for dilute mixtures. Hence, the reactant reservoir can pull the system away from the chemical equilibrium line, generically resulting in harvested chemical power in the case of reactant cycles. 

\begin{figure}[tb]
\centering
\includegraphics[width=0.85\linewidth]{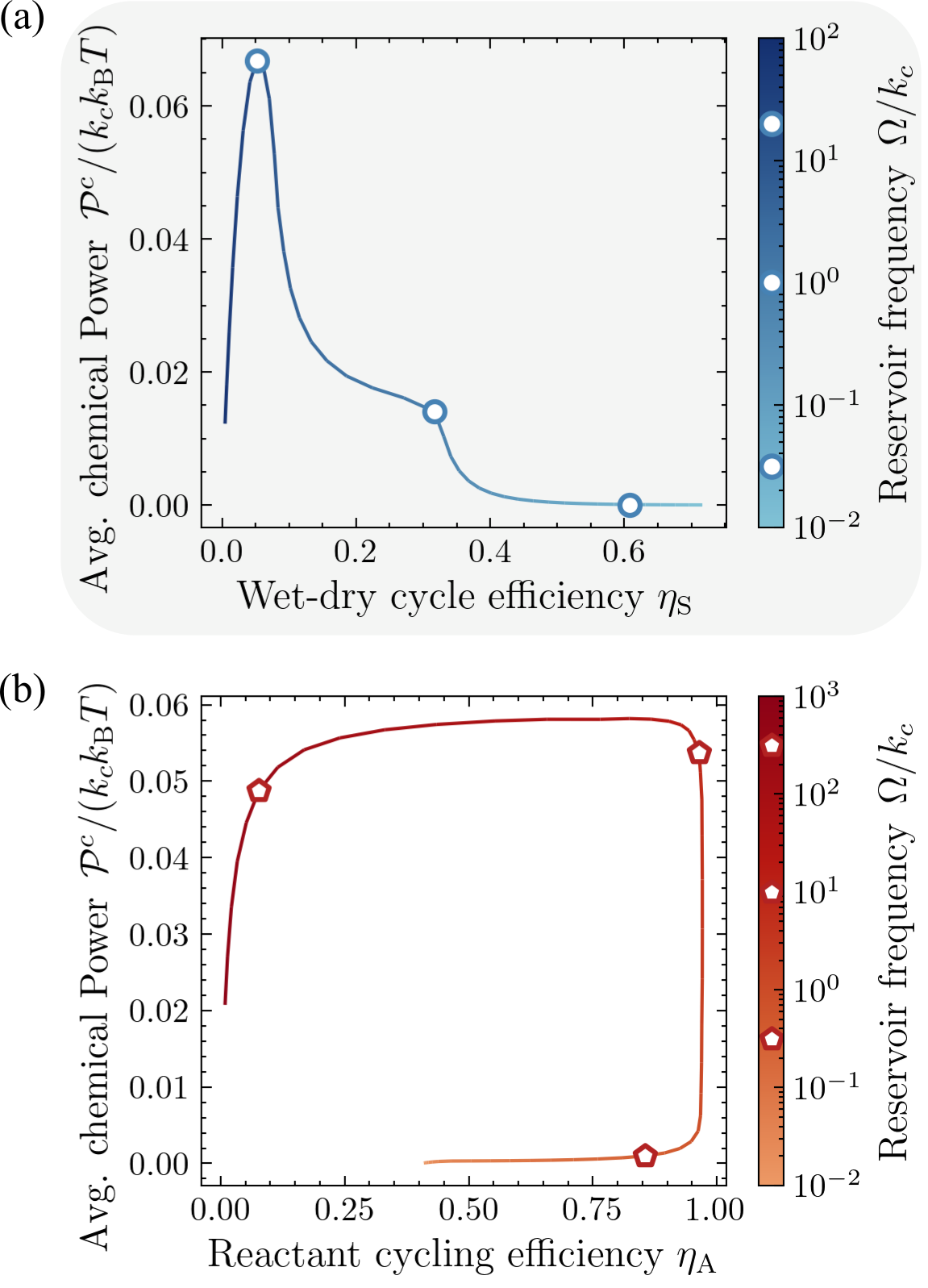}
\caption{\textbf{Wet-Dry and reactant cycles behave like different thermodynamic heat engines :} 
Power versus efficiency plots for reservoir exchange rate ($k_r/k_c = 10$) illustrate distinct behaviors of the two cycle types. (a) The wet-dry cycle exhibits characteristics similar to a Carnot engine, where high chemical power corresponds to low efficiency, and vice versa.
(b) In contrast, the reactant cycle resembles a stochastic heat engine, demonstrating a regime where both power and efficiency are high simultaneously. }
\label{fig:power_efficiency}
\end{figure}

\begin{figure*}[tb]
    \centering
    \makebox[\textwidth]{\includegraphics[width=\textwidth]{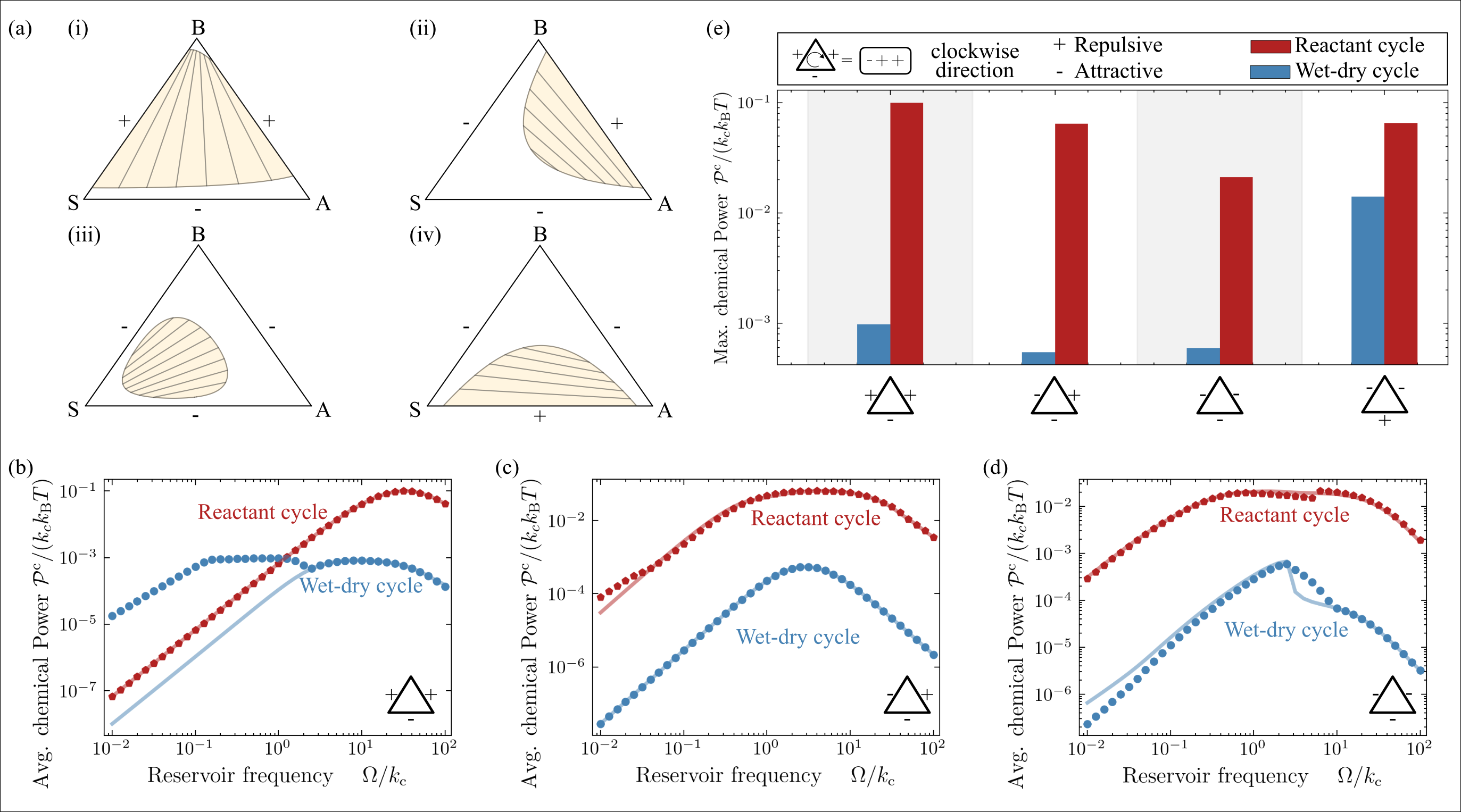}}
   \caption{
    \textbf{Effects of different interactions on harvesting chemical power.}
    (a)
    Four different cases of interactions are considered, reflected in phase diagrams of varying shapes. $-$/$+$ indicates attractive or repulsive interactions between the neighboring components in the triangular phase diagram. 
    For example, in the first diagram (clockwise reading: $`-++'$; see inset in (e)),  the solvent attracts each other $A$, while the solvent and $B$ and $A$ and $B$ repel each other, respectively. 
    (b,c,d)  
    Average chemical power $\mathcal{P}^c/(k_ck_\text{B}T)$ as a function of reservoir frequency $\Omega/k_c$ for the two cycle types (wet-dry cycles, reactant cycles) and the different interactions shown in (a), except the case $`+--'$ that is demonstrated in Fig.~\ref{fig:kinetic_rate_variation}(a,b).
    The solid transparent lines depict the mixed reference system (Eq.~\eqref{eq:phitotdot}).
    (e) Comparison of maximal chemical power between the different types of cycles, indicating that reactant cycles perform better in terms of harvesting power.}    \label{fig:diff_int_sys}
\end{figure*}

\subsection{System behaves similar to heat engines}\label{System_mimics_properties_of_different_heat_engines}

The wet-dry cycles exhibit characteristics similar to those of a Carnot engine, whereas reactant cycles display certain behaviors characteristic of a stochastic heat engine. Each thermodynamic engine exhibits specific trade-offs between power output and efficiency. 
By comparing our system to thermodynamic heat engines, we can quantify how effectively the considered cycling systems transfer energy and identify optimal operating conditions.
To this end, we depict efficiency $\eta$ for harvested chemical power using Eq.~\eqref{eq:sys_eff}. We characterize output power $\mathcal{P}^c$ versus efficiency $\eta$ for the reacting mixtures subject to wet-dry or reactant cycles. 

For wet-dry cycling, we observe that the system achieves high power output, but at the cost of low(er) efficiency (Fig.~\ref{fig:power_efficiency}(a)).
Conversely, for high efficiency, the power is lower. 
This is similar to a Carnot engine. 
To achieve maximum efficiency for a Carnot cycle, the engine must operate quasi-statically, which requires an infinitely slow cycle, resulting in zero power output. 
Conversely, operating the Carnot engine at finite speeds enables finite power generation, which comes at the cost of reduced efficiency. 
The reactant cycle, on the other hand, is similar to stochastic heat engines. 
The common property is that there is a frequency (see color bar of Fig.~\ref{fig:power_efficiency}(b)) at which the efficiency and power are maximal simultaneously.

\subsection{Effects of different interactions}\label{Effects_of_different_interactions}
To understand the role of phase separation in harvesting chemical power, we compare the harvested power of \(\mathcal{P}^c/(k_ck_\text{B}T)\) (Eq.~\eqref{eq:chemical_power}) in a phase-separating mixture to the corresponding mixed reference system (Eq.~\eqref{eq:phitotdot}).
We further vary the interactions between all three components, $A$, $B$, and the solvent in the reacting mixture. 
For simplicity, we discuss four systems with very different interactions for which we denote attraction/repulsion between the respective components as -/+, respectively. The corresponding phase diagrams are shown in Fig.~\ref{fig:diff_int_sys}(a). Thus, the four systems are abbreviated as: `- + +', `- - +', `- - -', and `+ - -'. For example, the `- + +' interactions mean that the solvent and $A$ attract each other, while the solvent and $B$ and $A$ and $B$ repel each other, respectively.
For a fair comparison between both reactant and wet-dry cycles, both reservoirs were oscillated around the system's mean thermodynamic point (defined as \(\mu_i^r(t) = \left<\mu_i^r\right>\)) with a common amplitude of cycle $\hat\mu_i$. 
We present the average chemical power for different reservoir cycles as we vary the cycling frequency; see Fig.~\ref{fig:diff_int_sys}(b-d) for a fixed reservoir kinetic rate of $k_r/k_c = 10$. 
We note that the power output for the system with `+ - -' interactions, also commonly known as Homotypic phase separation,  can be found above in Fig.~\ref{fig:kinetic_rate_variation}(a,b).

For all cases of interactions, we find that there is a maximum in chemical power \(\mathcal{P}^c_*\) that can be harvested from the reservoir, as shown in Fig.~\ref{fig:diff_int_sys}(e). 
Surprisingly, the maximal chemical power \(\mathcal{P}^c_*/(k_ck_\text{B}T)\) remains of comparable magnitude across different systems undergoing reactant cycle, whereas the maximum power by wet-dry cycle is notably low for the first three cases, except for the system with $A$-Solvent phobic `+ - -' interactions, as shown in Fig.~\ref{fig:diff_int_sys}(e).
The average chemical power by reactant cycles is always beneficial at all cycling frequencies for systems with interactions $A$-$B$ phobic  `- - +', (corresponding to Heterotypic phase separation) and hydrophilic  `- - -' ($A$-$S$ and $B$-$S$ likes to mix with solvent), see Fig.~\ref{fig:diff_int_sys}(c, d).
However, in the $B$-$S$ and $A$-$B$ phobic `- + +' interacting system, we see that the wet-dry cycle outperforms higher chemical power for lower cycling frequencies. At an intermediate frequency, we see a switch after which the reactant cycle yields higher chemical power.

Finally, we discuss how chemical power is affected when a mixture phase separates (scattered symbols) compared to a mixed system (solid lines), as shown in Fig.~\ref{fig:diff_int_sys}(b-d). 
At high frequencies, the system becomes unresponsive to changes by the reservoir, and the system's volume fraction trajectories become smaller, never entering the binodal region. Therefore, the chemical power is the same for mixed and phase-separating systems at high frequencies. On the contrary, for low frequencies, the harvested power can show an interesting behavior. For example, consider the case (i) `- + +' interactions, Fig.~\ref{fig:diff_int_sys}(b), where the average chemical power $\mathcal{P}^c$ in the presence of phase separation exhibits a local and a global maximum for the wet-dry cycle. We also see that phase separation enhances chemical work by two orders of magnitude. Due to phase separation, the free energy of the system is reduced, allowing more of the chemical power of the reservoir to be harvested. Therefore, we see that phase separation can enhance chemical power.

\section{Conclusion}\label{conclusion}

In this work, we present a thermodynamic framework for chemically reacting, non-dilute mixtures that exchange particles with an external reservoir undergoing periodic cycles. 
The cycling reservoir drives the system away from chemical equilibrium, making the mixture perform chemical work. 
This chemical work can be harvested and used to run other chemical processes or even to perform mechanical work.
We consider two types of cycling reservoirs: 
(i) A reactant reservoir, which exchanges periodically components of the mixture that participate in the chemical reaction, thereby generating reactant cycles.
(ii) A solvent reservoir, which exchanges a non-reacting, solvent component of the mixture, thereby generating wet-dry cycles. 
An important parameter governing the chemical power output from the reacting mixture is the cycling frequency $\Omega$ of the cycling external reservoir.

A key finding for both types of cycles is that a distinct cycling frequency $\Omega_*$ (resonance frequency) at which the power output of the chemically reacting mixture is maximal. For frequencies below and above $\Omega_*$, the power output decreases. 
For slow cycles of the reservoir, the mixture can relax back to thermodynamic equilibrium much faster than any changes occur in the reservoir. 
For fast cycles of the reservoir, the system is too slow to respond to the changes made by the reservoir.
At resonance frequency, the mixture has sufficient time to react to the external reservoir and can be driven away from chemical equilibrium and maximize chemical power $\mathcal{P}^c_*$. 

The distinct resonance frequencies for different reservoir flux rates can act as a selection mechanism, favoring the specific reaction pathways in prebiotic puddles of many molecules. A particularly important example is the selection of sequence patterns of RNA, where dense and phase-separated environments dictate the oligomerization pathways~\cite{haugerudTheorySequenceSelection2024,bartolucciInterplayBiomolecularAssembly2024c}, and cyclical phase transitions have been shown to select sequences~\cite{bartolucci2023sequence}.

We compare our chemically reacting system to different classes of thermodynamic heat engines.
We find that wet–dry cycles exhibit features analogous to a Carnot engine, whereas reactant cycles resemble stochastic heat engines.
In the limit of low-amplitude cycling ($\hat{\mu}_i/k_\text{B}T \ll 1$), we show that the dynamics of the reacting mixture coupled to a reservoir can be approximated by two coupled, driven harmonic oscillators that elongate and contract within the mixture’s thermodynamic phase diagram.
This approximation enables us to unravel how the harvested power depends on frequency and the rate coefficient of the chemical reactions. 

\begin{figure}[tb]
    \centering
    \includegraphics[width=0.9\linewidth]{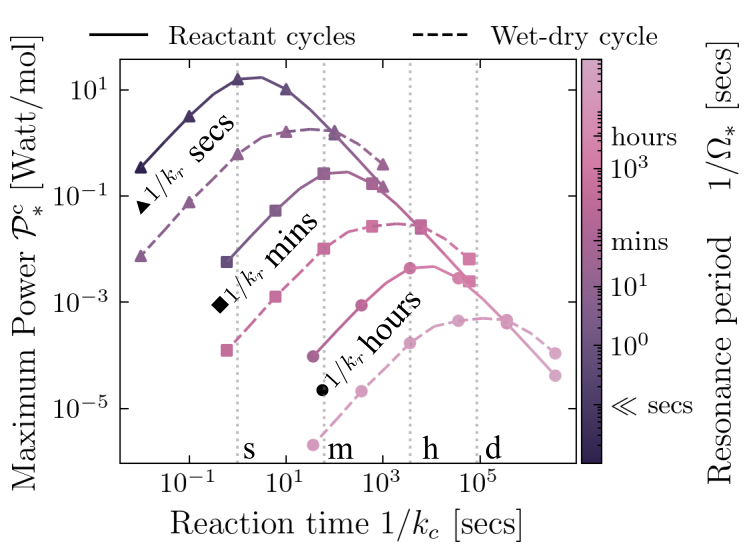}
    \caption{\textbf{Harvesting power from systems undergoing wet-dry and reactant cycles.} The power output by chemically reacting system in presence of cyclic reservoir is shown in real units. There exists an optimal speed of chemical reaction that enhances the maximum chemical power $\mathcal{P}_*^c$. Different reservoir timescales produces different power output. Power output from the system can be increased by making the exchange from the reservoir faster.
    For a system with a fixed reaction rate $k_c$, the maximal harvest is achieved when the cycle frequency is about the reaction rate ($\Omega\simeq k_c$), suggesting the selection of specific chemical processes.
    }
    \label{fig:conclusion_real_units}
\end{figure}

Our finding of non-dilute mixtures capable of efficiently harvesting chemical power from a cycling environment suggests the possibility of building complex molecular assemblies through cycling reservoirs, such as wet-dry cycles. 
This possibility is particularly relevant to the molecular origin of life, where there was no metabolic machinery to efficiently harvest energy from the surroundings. 
We found a maximal power harvested from the reservoir up to a few $10$ W/mol; see Fig.~\ref{fig:conclusion_real_units}.
Although the harvested power from the reservoir is smaller than that of the highly developed molecular motors in extant life (HMM and kinesin molecular motors are both around a few kW/mol~\cite{carter2005mechanics}), our results highlight that cyclic non-equilibrium environments on the early Earth could have provided the necessary kick-start for the evolution towards more high-energy molecules and molecular assemblies if the cycling frequency is around the resonance frequency.

Strikingly, non-mixtures with different chemical reactions (characterized by their reaction rate coefficient) harvest different amounts of power from the reservoir. 
There is an optimal reaction rate 
for which the harvested power is maximal. 
We speculate whether this has biased the selection to specific chemcial processes. Roughly speaking, maximal harvest is achieved when the reaction rate is about the cycle frequency $\Omega \simeq k_c$ (Fig.~\ref{fig:conclusion_real_units}).
Interestingly, template-directed primer-extension rates at prebiotic conditions are in the range $k_c\simeq 1/\text{day}$~\cite{kim2020model, prywes2016nonenzymatic,caimi2025high} -- a scenario that can be well sustained by wet-dry cycles stemming from the day-night rhythm at early Earth.

\begin{acknowledgements}

C.\ Weber and P.\ Jaiswal thank the University of Augsburg for financial support via the research program ``Forschungspotentiale besser nutzen!''. 
We thank D.\ Braun for the discussion on wet-dry cycles and their relevance to the molecular origin of Life. 
C.\ Weber thanks the TRR 392: Molecular evolution in prebiotic environments
(Project number 521256690) for support. 

\end{acknowledgements}

\appendix

\section{Gibbs free energy density and chemical potential for an interacting mixture}\label{app:gibbs_free_energy}

In the Gibbs ensemble, the system's chemical potential is defined as, 
\begin{align}\label{chem_po_appendix_1}
    \mu_i &= \frac{\partial {G}}{\partial N_i}\bigg|_{T, p,N_{j\neq i}} \, ,
\end{align}
where \({G}\) is the Gibbs free energy, which is defined in terms of the free energy density as,
\begin{align}
{G}/V = g(\{\bar\phi_i\},T,p) = \sum_{i=0}^M \frac{\omega_i\bar\phi_i}{\nu_i}+ k_{\text{B}}T\sum_{i=0}^M \frac{\bar\phi_i}{\nu_i}\log\bar\phi_i\nonumber\\+
\frac{1}{2}\sum_{i,j=0}^M\chi_{ij}\bar\phi_i\bar\phi_j+ p \, , 
\end{align}
where $\chi_{ij}$ denotes the interaction parameter between components $i$ and $j$, and $\omega_i$  is the internal energy. Using, Eq.~\eqref{chem_po_appendix_1} the chemical potential for each component $i$ thus becomes
\begin{align}
    \mu_i &= \omega_i 
        + k_\text{B}T\left(\log \bar\phi_i + 1\right) 
        + \frac{\nu_i}{\nu_\circ} \sum_{j=0}^M \chi_{ij} \bar\phi_j \nonumber \\
    &\quad   + \nu_i (p - \lambda) \nonumber \, , \\[0.5em]
    \text{where} \quad
    \lambda &= k_\text{B}T \sum_{i=0}^M \frac{\bar\phi_i}{\nu_i} 
        + \frac{1}{2\nu_\circ} \sum_{i=0}^M \sum_{j=0}^M \chi_{ij} \bar\phi_i \bar\phi_j\,.
\end{align}

\section{Harvesting chemical work from the energy dissipated by external reservoir}\label{energy_dissipation_and_work}

In this section, we derive the different contributions to the entropy production in the system and define the efficiency of the system for the reservoir cycles.
The model consists of two parts: the system and the reservoir.
Reversible work is done on the latter part in order to change the chemical potentials.
Both the system with free energy $F^s$ and the reservoir with free energy $F^r$  are in contact with a heat bath that maintains the temperature at $T$.
The total entropy production of the system and the reservoir is therefore given by
\begin{align}
dS^{tot} =
    - \frac{1}{T} dF^s 
    - \frac{1}{T} \left(  dF^r -  d\mathcal{W}^r \right).
\end{align}
After the transient, the system and reservoir state are periodic with a period $\tau=1/\Omega$, which means that
\begin{align}
\Delta F^r =
\lim_{t_0\rightarrow \infty} \int \limits_{t_0}^{t_0 + 1/\Omega} dF^r
 = F^r(t_0 + \tau) -  F^r(t_0) = 0 \, ,
\label{F_integral}
\end{align}
where $t_0$ is the initial time of the period.
Similarly, the free-energy difference of the system is 
$\Delta F^s = 0$.
Therefore, the entropy change over a period is
\begin{align}
\Delta S^{tot}
= \lim_{t_0\rightarrow \infty} \int \limits_{t_0}^{t_0 + 1/\Omega} dS^{tot}
= \frac{1}{T}\mathcal{W}^r,
\label{dissipated_heat}
\end{align}
where
\begin{align}
\mathcal{W}^r = \lim_{t_0\rightarrow \infty} \int \limits_{t_0}^{t_0 + 1/\Omega} d\mathcal{W}^r,
\end{align}
is the work done on the particle reservoir over one period.
Eq.~\eqref{dissipated_heat} states that, integrated over a period,
the energy added to the reservoir in the form of work ($\mathcal{W}^r$)
is equal to the total dissipated heat ($T \Delta S^{tot}$).

The state of the system changes due to chemical reactions and due to particle exchange with the reservoir.
The state of the reservoir change due to the particle exchange as well, and due to the work done on the reservoir.
The free-energy differences are therefore
\begin{align}
dF^s &= \sum_i \mu_i^s dN_i^s,
\\
dF^r &= \sum_i \mu_i^r dN_i^r + d\mathcal{W}^r .
\label{dFr}
\end{align}
Using that $dN_i^s = dN_i^c - dN_i^r$,
where $dN_i^c$ is the change due to chemical reactions,
the entropy change is therefor equal to
\begin{align}
dS^{tot}
=
 - \sum_i \frac{1}{T} \mu_i^s dN_i^c
 - \sum_i \frac{1}{T} (\mu_i^r - \mu_i^s) dN_i^r \, ,
\end{align}
which shows that the entropy change has two contributions:
the entropy change due to the dissipation in the chemical reactions (first term on the right-hand side),
and the entropy change due to the dissipation from the particle exchange between the system and particle reservoir (second term on the right-hand side).
We call the heat dissipated in the chemical reactions over one period
(minus the integral of first contribution times the temperature)
the chemical work done by the system:
\begin{align}
\mathcal{W}^c
=&
 \lim_{t_0\rightarrow \infty} \int \limits_{t_0}^{t_0 + 1/\Omega} 
  \sum_i \mu_i^s \text{d}N_i^c\nonumber\\
=&
 \lim_{t_0\rightarrow \infty} \int \limits_{t_0}^{t_0 + 1/\Omega}  \text{d}t~
   \sum_\alpha \Delta \mu_\alpha V r_\alpha \, ,
\end{align}
where we used
$\sum_i \mu_i^s \text{d}N_i^c  = \sum_\alpha \Delta \mu_\alpha V r_\alpha\Delta t $, here $r_\alpha$, is the reaction flux for a chemical reaction. In case of a volume conserving $A\rightleftharpoons B$ reaction we have $r_\alpha \equiv r$ given in Eq.~\eqref{eq:reaction_rate2}.
Taking the $\text{lim} ~ t_0  \rightarrow \infty$ such that the system and reservoir are periodic with time period $1/\Omega$.

The work done on the cycling the reservoir over a period can be expressed in the system parameters by integrating Eq.~\eqref{dFr} over a period and using Eq.~\eqref{F_integral}, which gives
\begin{align}
\mathcal{W}^r
=&  -\lim_{t_0\rightarrow \infty}  \int \limits_{t_0}^{t_0 + 1/\Omega} \sum_i \mu_i^r dN_i^r
\nonumber \\
=&  \lim_{t_0\rightarrow \infty} \int \limits_{t_0}^{t_0 + 1/\Omega} dt 
        \sum_i \mu_i^r V h_i / \nu_i \, ,
\end{align}
where we used that $-dN_i^r = dt \, V h_i / \nu_i$ is the particle flux of species $i$ from the reservoir into the system.

In the case of an inhomogeneous system, the work done on the system becomes
\begin{align}
\mathcal{W}^r
=&  \lim_{t_0\rightarrow \infty} 
    \int  \limits_{t_0}^{t_0 + 1/\Omega} dt 
    \int \limits_{\partial \Phi_r} d^2 \bm{x}_b
        \sum_i \mu_i^r \tilde{h}_i(\bm{x}_b) / \nu_i\, ,
\end{align}
where $\tilde{h}_i(\bm{x}_b)$ is the reservoir flux per unit area.
The chemical work becomes
\begin{align}
\mathcal{W}^c
=
 \lim_{t_0\rightarrow \infty}
 \int  \limits_{t_0}^{t_0 + 1/\Omega}  dt~
 \int \limits_{V(t)} d^3\bm{x}
   \sum_\alpha \Delta \mu_\alpha R_\alpha\, .
\end{align}

We define the efficiency of the system as the ratio of the chemical work and the total 
dissipated heat ($T \Delta S^{tot} = \mathcal{W}^r$):
\begin{align}
\eta_s
=& \frac{\mathcal{W}^c}{T \Delta S^{tot}}
=
\lim_{t_0\rightarrow \infty} 
\frac{
 \int \limits_{t_0}^{t_0 + 1/\Omega} dt~ \sum_\alpha \Delta \mu_\alpha R_\alpha
}{
\int \limits_{t_0}^{t_0 + 1/\Omega} dt~ \sum_i  \mu^r_i V h_i / \nu_i
} \,.
\end{align}
 
\section{Orbit solutions for small reservoir cycle amplitudes and definition of damping matrix  $\uuline{\mathbf{\Gamma}}$}
\label{analytical_orbits_for_small_amplitude}

The deviation of the system’s average volume fraction around the mean thermodynamic equilibrium point $\{\bar\phi_i^*\}$ decreases when the reservoir is cycled with a small amplitude. 
As shown as an example case of the reactant cycle in Fig.~\ref{fig:amplitude_change}(a), the compositional orbits contract toward the mean thermodynamic equilibrium point (black star) for an amplitude $\hat\mu_A/2$. 
We can define the systems average composition as $\{\bar\phi_i^* +\delta\bar\phi_i\}$, where $\delta\bar\phi_i$ is the deviations in the system average composition from the mean thermodynamic equilibrium point.
The systems kinetics is governed by a volume conserving chemical reaction $A \rightleftharpoons B$, and exchange from a reservoir $x$, which can either exchange reactant molecules ($x\equiv A$) or solvent molecules ($x\equiv S)$. 
The deviation in chemical $\delta r_i$ and the reservoir flux $\delta h_{i,x}$ around any arbitrary point $\{\bar\phi_i\}$ is given as,
\begin{align}
    \delta r_i &=r_i(\{\bar\phi_i +\delta\bar\phi_i\}) -r_i(\{\bar\phi_i\}) \,,\nonumber\\[0.5em]
    \delta h_{i,x} & = h_{i,x}(\{\bar\phi_i + \delta\bar\phi_i\}) - h_{i,x}(\{\bar\phi_i\})\,.
\end{align}
Choosing this point to be the mean thermodynamic equilibrium point $\{\bar\phi_i^*\}$, the flux in chemical and reservoir at $\{\bar\phi_i^*\}$ becomes zero, i.e, $r_i(\{\bar\phi_i^*\}) = 0$ and $h_{i,x}(\{\bar\phi_i^*\}) = 0$.

In the limits of small amplitude ($\hat\mu_i\ll k_\text{B}T$), the deviations in systems average  $\{\delta\bar\phi_i\}$ becomes smaller.
Therefore, we evaluate the flux in system's compositional deviation $\{\delta\bar\phi_i\}$ from its mean equilibrium as deviation in chemical flux as,
\begin{gather}
    \partial_t \delta\bar\phi_i = \delta r_i + \delta h_{i,x} = \delta r_i +(\delta^r_{i,x}- \bar\phi_i^*)\delta h_x\,,\nonumber\\[1em]
    \partial_t \delta V = V^*\delta h_x\,,
    \label{eq:lin_tim_evo}
\end{gather}
where $\delta h_{i,x}$ are the deviations in the reservoir flux around the mean thermodynamic point for component ($i$) due to a reservoir ($x$). Similarly, $\delta r_i$ is the deviation in chemical flux around $\{\bar\phi_i^*\}$ and $V^*$ is the systems volume at $\{\bar\phi_i^*\}$.

To obtain an expression for $\delta r_i$ and $\delta h_{i,x}$, we first need to Taylor expand the chemical potential in terms of deviations around the mean thermodynamic equilibrium as,
\begin{align}
    \mu_i(\{\bar\phi_i^* + \delta\bar\phi_i\}) &\simeq \mu_i(\{\bar\phi_i^*\})   \nonumber\\[0.5em]
    &\quad + \sum_j \left.\frac{\partial\mu_i}{\partial \bar\phi_j}\right|_{\{\bar\phi_i^*\}}\delta \bar\phi_j + \mathcal{O}\left(\delta \bar\phi_j^2\right)\,.
    \label{eq:lin_chem_pot}
\end{align}
Here, we denote the zeroth-order term as \(\mu_i(\{\bar\phi_i^*\}) = \mu_i^*\), i.e., the chemical potential at the mean thermodynamic equilibrium $\{\bar\phi_i^*\}$. 
The first-order linear coefficient is
\begin{equation}
 \left.\frac{\partial\mu_i}{\partial \bar\phi_j}\right|_{\{\bar\phi_i^*\}} =\mu_{ij}^* \, .
\end{equation}
These derivatives of the chemical potentials are evaluated at the mean thermodynamic equilibrium. 
We obtain deviations in chemical flux $\delta r_i$ by substituting the linearized chemical potential Eq.~\eqref{eq:lin_chem_pot} in Eq.~\eqref{eq:reaction_rate2}  as, 
\begin{align}
\label{eq:linear_chemical_flux}
    \delta r = k_c^*\left[\left\{\frac{\mu_{AA}^* - \mu_{BA}^*}{k_\text{B}T}\right\}\delta\bar\phi_A+\left\{\frac{\mu_{AB}^* - \mu_{BB}^*}{k_\text{B}T}\right\}\delta\bar\phi_B\right]\,.
\end{align}
where $\delta r_A =\, -\delta r_B =\, -\delta r$. 
We use the condition of chemical equilibrium at $\mu_A^* = \mu_B^*$, and the reaction rate coefficient gets multiplied by a constant factor as $k_c^* = k_c\exp(\mu_A^*/k_\text{B}T) = k_c\exp(\mu_B^*/k_\text{B}T)$. 

Similarly, we obtain deviation in volume fraction reservoir flux $\delta h_x$ by substituting the linearized chemical potential Eq.~\eqref{eq:lin_chem_pot} in Eq.~\eqref{eq:boundary_flux_integrated}  as,
\begin{equation}
\label{eq:linear_volume_flux}
    \delta h_x=\, h_{\text{driving},x}(t) + h_{\text{relax},x}(\{\delta\bar\phi_i\})\,,
\end{equation}
where  $h_{\text{relax},x}$ relaxes the system towards mean thermodynamic equilibrium, and $h_{\text{driving},x}$ drives the system away from it. Both $h_{\text{relax},x}$ and $h_{\text{driving},x}$ is given as
\begin{gather}
    h_{\text{relax},x} = -k_r^*\frac{\nu}{V^*}\left(\frac{\mu_{xA}^*\delta\bar\phi_A + \mu_{xB}^*\delta\bar\phi_B}{k_\text{B}T}\right)\,,\nonumber \\[0.5em]
    h_{\text{driving},x} = k_r^*\frac{\nu}{V^*}\left(\frac{\hat\mu_x\sin(2\pi\Omega t)}{k_\text{B}T}\right)\,.
\label{eq:linear_volume_flux_2}
\end{gather}
We have used the condition of reservoir equilibrium  as $\left<\mu_x^r\right> = \mu_x^*$, and the reservoir exchange rate coefficient gets multiplied by a constant factor $k_r^* = k_r\exp(\left<\mu_x^r\right>/k_\text{B}T) = k_r\exp(\mu_x^*/k_\text{B}T)$. Both $\delta r_i$ and $\delta h_x$ are linearized fluxes in deviations $\{\delta\bar\phi_i\}$. 
The net exchange flux vector $\bm{h}_{r}$ from the reservoir around the mean thermodynamic point is given as
\begin{equation}
\label{eq:vector_exchange_div}
    \bm{h}_{r} = \delta h_x(\delta_{A,x}^r - \bar\phi_A^*,\,\delta_{B,x}^r - \bar\phi_B^*)\,.
\end{equation}
We use Eq.~\eqref{eq:linear_volume_flux} and Eq.~\eqref{eq:linear_volume_flux_2} to write the fluxes for the relaxation to mean thermodynamic equilibrium $\bm{h}_{\text{relax}}$ and external driving away from the mean equilibrium $\bm{h}_{\text{driving}}$ as,
\begin{gather}
    \bm{h}_{\text{relax}} \equiv\, h_{\text{relax},x}(\delta_{A,x}^r - \bar\phi_A^*,\,\delta_{B,x}^r - \bar\phi_B^*)\nonumber \, , \\[0.5em]
    \bm{h}_{\text{driving}}\equiv\,h_{\text{driving},x}(\delta_{A,x}^r - \bar\phi_A^*,\,\delta_{B,x}^r - \bar\phi_B^*) \, . 
\label{eq:fluxes_in_relax_driving}
\end{gather}
The relaxation flux vector $\bm{h}_{\text{relax},x}$ is shown in black arrow in Fig.~\ref{fig:amplitude_change}(b).

Substituting Eq.~\eqref{eq:linear_chemical_flux} and Eq.~\eqref{eq:linear_volume_flux} into Eq.~\eqref{eq:lin_tim_evo}, as a result we obtain,
\begin{equation}
\label{eq:lin_time_evo_matrix}
    \partial_t\delta\bm{\Phi} = -\mat{\Gamma}\cdot\delta\bm{\Phi} + \mathbf{R}(t),\,\,\,\,\text{with}\,\,\,\delta\bm{\Phi}=
    \begin{pmatrix}
        \delta\bar\phi_A\\[0.5em]
        \delta\bar\phi_B
    \end{pmatrix}.
\end{equation}
Here, the coefficient of the first term $\mat{\Gamma}$ is a damping term which splits into contributions from chemical reaction $\mat{\bm{\Gamma}}_c$ and reservoir $\mat{\bm{\Gamma}}_{r,x}$ as, 
\begin{equation}
\mat{\Gamma} = \mat{\Gamma}_c + \mat{\Gamma}_{r,x}.
\end{equation}
The elements of the damping coefficient matrix form chemical reactions can be written as, 
\begin{equation}
     \mat{\bm{\Gamma}}_c =\frac{k_c^*}{k_BT}
    \begin{pmatrix}
         \mu_{AA}^* - \mu_{BA}^* & \quad\mu_{AB}^* - \mu_{BB}^* \\[1em]
         \mu_{AB}^* - \mu_{AA}^* &\quad \mu_{BB}^* - \mu_{AB}^*
    \end{pmatrix}
\end{equation}
and the elements of damping coefficient matrix form reservoir $\mat{\Gamma}_{r,x}$, depends on the type of reservoir considered. i.e., reactant ($x = A$) or solvent ($x = S$) making the matrix as, 
\begin{equation}
    \mat{\bm{\Gamma}}_{r,x} = \frac{\nu k_r^*}{V^*k_\text{B}T}
    \begin{pmatrix}
        (\delta_{A,x}^r - \bar\phi_A^*)\mu_{xA}^* & \,\,(\delta_{A,x}^r - \bar\phi_A^*)\mu_{xB}^*\\[1.0em]
        (\delta_{B,x}^r - \bar\phi_B^*)\mu_{xA}^* &\,\, (\delta_{B,x}^r - \bar\phi_B^*)\mu_{xB}^*
    \end{pmatrix} \, . 
\end{equation}
The second term in Eq.~\eqref{eq:lin_time_evo_matrix} prevents the system from relaxing towards the mean thermodynamic equilibrium. It depends on the cycling amplitude of the reservoir $\hat\mu_x$ as,
\begin{equation}
\label{eq:driving_vec}
    \mathbf{R}(t) = \frac{\hat\mu_x^*}{k_\text{B}T}\sin(2\pi\Omega t)
    \begin{pmatrix}
        \delta_{A,x}^r - \bar\phi_A^*\\[1.5em]
        \delta_{B,x}^r - \bar\phi_B^*
    \end{pmatrix} \, , 
\end{equation}
with $\hat\mu_x^* = {k_r^*\nu}/({V^*})\hat\mu_x$.
Here, $\mathbf{R}(t)$ is a time-dependent driving term coming from the external reservoir.\par

The solutions to the deviations of systems' average volume fractions around mean thermodynamic equilibrium are,  
\begin{equation}
\label{analytics:sol_2}
    \delta \bm{\Phi}(t)=\, \dfrac{\hat\mu_x^*}{k_\text{B}T}\int_{t_{0}}^t e^{-\mat{\mathbf{\Gamma}}(t-t')}\cdot\begin{pmatrix}
        \delta^r_{A,x} -\bar\phi_A^*\\[1em] 
        \delta^r_{B,x}-\bar\phi_B^*
    \end{pmatrix}\sin(2\pi\Omega t')\text{d}t' \, .
\end{equation}
The damping matrix $\mat{\mathbf{\Gamma}}$ can be diagonalized in terms of its eigenvalues $\lambda_{\mp}$ as,
\begin{equation}
\label{eq:diagonalise}
    \mat{\mathbf{\Gamma}} = \mat{\mathbf{Y}}\cdot
    \begin{pmatrix}
        \lambda_-&0\\[1em]
        0&\lambda_+
    \end{pmatrix}\cdot\mat{\mathbf{Y}}^{-1},\,\text{where}\,\,\mat{\mathbf{Y}}\cdot\mat{\mathbf{Y}}^{-1} = \mat{\mathbf{I}} \, . 
\end{equation}
The exponential in Eq.~\eqref{analytics:sol_2} can be expressed in terms of a power series of the damping matrix  $\mat{\mathbf{\Gamma}}$.  
Using Eq.~\eqref{eq:diagonalise}, we diagonalize each term in the power series as,
\begin{equation}
\label{analytics:sol_3}
    \delta\bm{\Phi}(t)=\, \frac{\hat\mu_x^*}{k_\text{B}T}\int^{t}_{t_{0}}\mat{\mathbf{Y}}\cdot\mat{\mathbf{P}}\cdot\mat{\mathbf{Y}}^{-1}\cdot\begin{pmatrix}
        \delta^r_{A,x}-\bar\phi_A^*\\[1em] \delta^r_{B,x}-\bar\phi_B^*
    \end{pmatrix}\sin(2\pi\Omega t^{'})\text{d}t^{'} \, , 
\end{equation}
where
\begin{equation}
\mat{\mathbf{P}} =
    \begin{pmatrix}
    e^{-\lambda_{-}(t-t^{'})}&0\\[8pt] 0 &e^{-\lambda_{+}(t-t^{'})}
    \end{pmatrix} \, . 
\end{equation}
Integrating the above Eq.~\eqref{analytics:sol_3}, we get an expression for deviation around the mean thermodynamic equilibrium point as,
\begin{equation}
\label{analytics:sol_4}
    \delta\bm{\Phi}(t) =\, \frac{\hat\mu_x^*}{k_\text{B}T}\mat{\mathbf{Y}}\cdot\mat{\mathbf{D}}(t)\cdot\mat{\mathbf{Y}}^{-1}\cdot\begin{pmatrix}
        \delta^r_{A,x}-\bar\phi_A^*\\[1em]
        \delta^r_{B,x}-\bar\phi_B^*
    \end{pmatrix},
\end{equation}
with 
\begin{equation}
    \mat{\mathbf{D}} = 
    \begin{pmatrix}
    \frac{\lambda_-\sin(\widetilde\Omega t) - \widetilde\Omega \cos(\widetilde\Omega t)}{\widetilde\Omega^2 + \lambda_-^2}&0\\[1em]0 &\frac{\lambda_+\sin(\widetilde\Omega t) - \widetilde\Omega \cos(\widetilde\Omega t)}{\widetilde\Omega^2 + \lambda_+^2}
    \end{pmatrix} \, . 
\end{equation}
Here, we redefine the angular frequency as $\widetilde\Omega = 2\pi\Omega$.
The full expression for the deviation from the mean thermodynamic equilibrium for each component $\delta\bar\phi_i$ is given as,
\begin{align}
\label{analytics:sol_5}
    \delta\bar\phi_i(t) = \frac{\hat\mu_x^{*}}{k_\text{B}T} \Bigg[ 
         \sin(\widetilde\Omega t)\left\{
            \frac{\lambda_-}{\widetilde\Omega^2+\lambda_-^2}a_i + \frac{\lambda_+}{\widetilde\Omega^2+\lambda_+^2}b_i
        \right\} \nonumber \\[1em]
        -  \cos(\widetilde\Omega t)\left\{
            \frac{\widetilde\Omega}{\widetilde\Omega^2+\lambda_-^2}a_i + \frac{\widetilde\Omega}{\widetilde\Omega^2 + \lambda_+^2}b_i
        \right\}
    \Bigg]
     \, . 
\end{align}
In Eq.~\eqref{analytics:sol_5}, the $a_i$ and $b_i$ (for $i\in\{A,B\}$) are constants define as, 
\begin{align}
    a_A &= y_{11}(y^{-1}_{11}(\delta^r_{A,x} - \bar\phi_A^*) + y^{-1}_{12}(\delta^r_{B,x}-\bar\phi_B^*))\,,\nonumber\\[0.5em] 
    b_A &= y_{12}(y^{-1}_{21}(\delta^r_{A,x} - \bar\phi_A^*) + y^{-1}_{22}(\delta^r_{B,x}-\bar\phi_B^*))\,,\\[0.5em]
    a_B &= y_{21}(y^{-1}_{11}(\delta^r_{A,x} - \bar\phi_A^*) + y^{-1}_{12}(\delta^r_{B,x}-\bar\phi_B^*))\,,\nonumber\\[0.5em]
    b_B &= y_{22}(y^{-1}_{21}(\delta^r_{A,x} - \bar\phi_A^*) + y^{-1}_{22}(\delta^r_{B,x}-\bar\phi_B^*))\,.\nonumber 
\end{align}
Here, $y_{ij}$ and $y^{-1}_{ij}$ are the elements of the matrix $\mat{\mathbf{Y}}$ and $\mat{\mathbf{Y}}^{-1}$, respectively.
These are constants that depend on the mean thermodynamic equilibrium. Interestingly, from Eq.~\eqref{analytics:sol_5}, we see that the deviations from the mean thermodynamic equilibrium point scale linearly with the cycling amplitude $\hat\mu_x$.

\section{Mobilities of orbital points in the phase diagram when subject to an effective spring network with spring constant $\uline{\mathbf{K}}$}
\label{app:mobilities_and_forces}
The dynamics of the reacting mixture, subject to a reservoir, correspond to two coupled and driven oscillators that elongate and retract in the mixture's thermodynamic phase diagram.
The equation of motion of a driven object of mass $m$ connected to a wall by a harmonic spring with spring constant $\mat{\mathbf{K}}$ is written as, 
\begin{equation}
    m\, \delta\ddot{\bm{\Phi}} + \mat{\mathbf{K}}\cdot\delta\bm{\Phi} + \mat{\mathbf{M}}^{-1}\cdot \delta\dot{\bm{\Phi}}= \widetilde{\mathbf{R}}(t)\,.
\end{equation}
In the overdamped limit, the acceleration of the system vanishes, making $\delta\ddot{\bm{\Phi}} = 0$, which results in:
\begin{equation}
\label{eq:spring-mass-system}
    \delta\dot{\bm{\Phi}} = -\mat{\mathbf{M}}\cdot\mat{\mathbf{K}} \cdot \delta\bm{\Phi} + \mat{\mathbf{M}}\cdot\widetilde{\mathbf{R}}(t)\,.
\end{equation}
In the above Eq.~\eqref{eq:spring-mass-system}, the first term is the damping flux that relaxes the system towards equilibrium.
This results from the product of the mobility matrix of the system $\mat{\mathbf{M}}$ with the force $\mathbf{F}$. 
A harmonic spring with deviation $\delta\bm{\Phi}$ from its rest length, results in a force $\mathbf{F} = -\mat{\mathbf{K}}\cdot\delta\bm{\Phi}$.
Similarly, expanding the equation of motion for a chemically reacting system with exchange from a reservoir around the mean thermodynamic equilibrium gives:
\begin{equation}
\label{eq:lin_time_evo_matrix_expanded} 
    \partial_t\delta\bm\Phi = \mat{\mathbf{M}}_c\cdot\mathbf{F}_c + \mat{\mathbf{M}}^x_{r}\cdot\mathbf{F}_{r,x} + \mathbf{R}(t) \, . 
\end{equation}
These equations suggest that the dynamics of the reacting mixture, subject to a reservoir, correspond to two coupled and driven oscillators elongating and retracting in the mixture's thermodynamic phase diagram as shown in Fig.~\ref{fig:amplitude_change}(c).
Note that $x = A$ indicates reactant cycles and $x = S$ wet-dry cycles. Moreover, the subscripts c and r label chemical reaction and reservoir-related quantities, respectively. 

The system (denoted by a white symbol) is attached to the chemical and reservoir equilibrium line with springs of spring constant matrices $\mat{\mathbf{K}}_c$ and $\mat{\mathbf{K}}_{r,x}$, respectively.
The thermodynamic forces form the chemical reaction and the reservoir are,
\begin{subequations}
 \label{eq:force} 
\begin{align}
    \mathbf{F}_c &= -\mat{\mathbf{K}}_c\cdot\delta\bm{\Phi}\, ,
    \\
\mathbf{F}_{r,x} &= -\mat{\mathbf{K}}_{r,x}\cdot\delta\bm{\Phi}\,.
\end{align}
\end{subequations}
We obtain the respective spring constant matrices from the inverse relation,
\begin{subequations}
\label{eq:inv_rel}
\begin{align}   \mat{\mathbf{K}}_c &= \left.(\mat{\mathbf{M}}_c)^{-1}\right|_{\{\bar\phi_i^*\}}\cdot\left.\mat{\mathbf{\Gamma}}_c\right|_{\{\bar\phi_i^*\}}\,,\\[0.5em]
    \mat{\mathbf{K}}_{r,x} &= \left.(\mat{\mathbf{M}}^x_{r})^{-1}\right|_{\{\bar\phi_i^*\}}\cdot\left.\mat{\mathbf{\Gamma}}_{r,x}\right|_{\{\bar\phi_i^*\}}\,,
\end{align}
\end{subequations}
with $\left.\mat {\mathbf{M}}_{c}\right|_{\{\bar\phi_i^*\}}$ and $\left.\mat {\mathbf{M}}^x_{r}\right|_{\{\bar\phi_i^*\}}$ are the mobility matrices for fluxes due to chemical reaction and exchange of molecules from reservoir ($x$), respectively, at the mean thermodynamic equilibrium point.\par

To obtain the mobilities and spring matrices, we start by looking at the fluxes for the system shown in Eq.~\eqref{eq:kinetic_evolve}, which can be written in a simplified way for a homogeneous system that undergoes chemical reactions $A\rightleftharpoons B$, and can exchange reactant ($A$) or solvent molecules ($S$) within the reservoir as,
\begin{equation}
\label{eq:homo_flux}
    \partial_t \bar\phi_i = r_i + h_{i,x}\,,\quad\text{and}\quad\partial_t V = V h_x\,.
\end{equation}
where $r_i$ is the volume fraction flux due to chemical reaction and $h_{i,x}$ is the volume fraction flux from the reservoir. Depending on the type of reservoir, either reactant ($x \equiv A$) or solvent ($x \equiv S$), we can write the reservoir flux as,
\begin{equation}
    h_{i,x} = (\delta^r_{i,x} - \bar\phi_i)h_x\,.
\end{equation}
The reaction flux vector $\bm{r}_c$ shown in orange arrow and exchange flux vector $\bm{h}_r$ shown in black arrow in Fig.~\ref{fig:orbit_shape_and_flux} is given as,

\begin{subequations}
\begin{align}
 \bm{r}_c&\equiv\,(r_A, r_B) = r(-1,1)\, , \\[0.5em]
\bm{h}_r&\equiv\,(h_A,h_B) = h_x(\delta_{A,x}^r - \bar\phi_A, \delta_{B,x}^r - \bar\phi_B)\,,
\end{align}
\end{subequations}
where the reaction flux $r$ is given in Eq.~\eqref{eq:reaction_rate2} (skipping the phase index $\alpha$ since we consider a homogeneous mixture in this section). The thermodynamic forces for bi-directional chemical reactions $A\rightleftharpoons B$ are given as:
\begin{equation}
\label{eq:thermo_force_cr}
    \Delta\mu_c^\leftharpoondown=-\Delta\mu_{c}^{\rightharpoonup} = \mu_A - \mu_B\,.
\end{equation}
The thermodynamic force due to exchange from the reservoir $x$ is read:
\begin{equation}
\label{eq:thermo_force_res}
    \Delta\mu_{x,r} = -(\mu_x^r - \mu_x) \, . 
\end{equation}
Around mean thermodynamic equilibrium, for small differences in thermodynamic force Eq.~\eqref{eq:thermo_force_cr} and ~\eqref{eq:thermo_force_res}, the fluxes in the system from linear response can be written as,
\begin{align}
    \partial_t\bar{\phi}_A &\simeq \frac{M_{A,c}}{k_\text{B}T}\Delta\mu_c^\rightharpoonup + \frac{M_{A,r}^{x}}{k_\text{B}T}\Delta\mu_{x,r}\,,\label{eq:lin_res_phi_and_V_1}\\[0.5em]
    \partial_t\bar{\phi}_B &\simeq \frac{M_{B,c}}{k_\text{B}T}\Delta\mu_c^\leftharpoondown + \frac{M_{B,r}^{x}}{k_\text{B}T}\Delta\mu_{x,r}\,,\label{eq:lin_res_phi_and_V_2}\\[0.5em]
    \partial_t V &\simeq \frac{M_{V,r}^{x}}{k_\text{B}T}\Delta \mu_{x,r}
    \, , \label{eq:lin_res_phi_and_V_3}
\end{align}
where the mobilities $M_{i,c}$ and $M_{i,r}^x$ are the respective mobilities for component $i=A,B$ due to fluxes from chemical reaction (c) or exchanges with the reservoir (r). 
The mobility from the reservoir depends on the reservoir type as $M_{i,r}^{x} = M_{x,r}(\delta_{i,x}^r- \bar\phi_i)$, where $M_{x,r} = \nu k_r^*/V$. 
A reactant reservoir, i.e, ($x\equiv A$), the mobilities for volume fraction fluxes in $A$ and $B$ are,
\begin{align}
    M_{A,r}^A = \dfrac{\nu k_r^*}{V}(\bar\phi_A-1)\,,\quad
    M_{B,r}^A =\frac{\nu k_r^*}{V}\bar\phi_B\, , 
\end{align}
and for a solvent reservoir, i.e., ($x\equiv S$), these coefficients are, 
\begin{align}
    M_{A,r}^S = \dfrac{\nu k_r}{V}\bar\phi_A\,,\quad
    M_{B,r}^S =  \frac{\nu k_r}{V}\bar\phi_B \, . 
\end{align}
For chemical reactions, the mobilities are,
\begin{equation}
    M_{A,c} = M_{B,c} = k_c^* \, . 
\end{equation}
Fluxes in Eq.~\eqref{eq:lin_res_phi_and_V_1}-\eqref{eq:lin_res_phi_and_V_3} are evaluated around the mean thermodynamic point $\{\bar\phi_i^*\}$ by expanding the forces around the mean thermodynamic equilibrium point.
The fluxes in deviation $\delta\bar\phi_i$ do not depend on the change in volume $\delta V$; however, the fluxes in volume deviations $\delta V$  depend on the deviations $\delta\bar\phi_i$ as shown in Eq.~\eqref{eq:lin_tim_evo} using Eq.~\eqref{eq:linear_volume_flux_2}.
From the inverse relationship given in Eq.~\eqref{eq:inv_rel}, we obtain the respective spring constant matrices for the springs shown in Fig. \ref{fig:amplitude_change}(c). 
The spring connecting the system to the chemical equilibrium line is given as,
\begin{equation}
\label{eq:mat_chem_spring_const}
    \mat{\mathbf{K}}_c = 
    \begin{pmatrix}
        \dfrac{\mu_{AA}^* - \mu_{BA}^*}{k_\text{B}T} & \dfrac{\mu_{AA}^* - \mu_{BA}^*}{k_\text{B}T}\\[1em]
        \dfrac{\mu_{AB}^* - \mu_{BB}^*}{k_\text{B}T} & \dfrac{\mu_{AB}^* - \mu_{BB}^*}{k_\text{B}T}
    \end{pmatrix}
\end{equation}
and the spring connecting the system to the reservoir equilibrium line is given as,
\begin{equation}
\label{eq:mat_resv_spring_const}
    \mat{\mathbf{K}}_{r,x} = 
    \begin{pmatrix}
        -\dfrac{\mu_{xA}^*}{k_\text{B}T} & -\dfrac{\mu_{xA}^*}{k_\text{B}T}\\[1em]
        -\dfrac{\mu_{xB}^*}{k_\text{B}T} & -\dfrac{\mu_{xB}^*}{k_\text{B}T}
    \end{pmatrix} \, . 
\end{equation}

\section{Phase separation in a non-dilute mixture can give rise to multiple local maxima of chemical power}
\label{local_global}

\begin{figure}[h]
    \centering
 \makebox[\textwidth]{\includegraphics[width=0.85\textwidth]{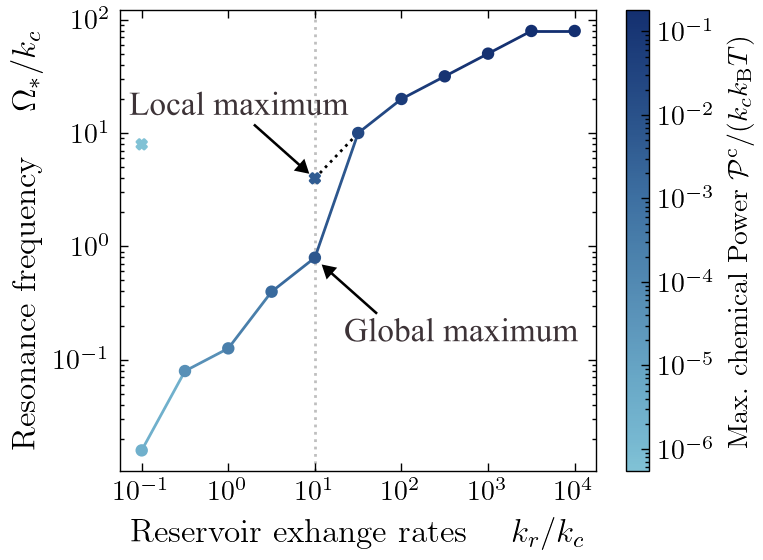}}   \caption{\textbf{Jump in the resonance frequency:} Resonance frequency that maximizes average chemical power $\mathcal{P}^c_*$ is shown for a system undergoing a wet-dry cycle with non-dilute interactions. At \(k_r/k_c = 10\), we see that there can be more than one frequency (local (cross symbol)/global (dot symbol)) around which chemical work gets optimized. A switch in the resonance frequency is observed at \(k_r/k_c =10\).}     \label{fig:lo_gl_max}
\end{figure}

\section{Power output by chemically reacting system for small amplitude}\label{analytical_power_output}

A system with chemical reactions can exhibit power output $\mathcal{P}^c$ when the external reservoir drives the system away from chemical equilibrium $\Delta \mu_c \neq 0$; see Eq.~\eqref{eq:chemical_power} for the expression. 
For reservoir cycles with small amplitude ($\hat\mu_i/k_\text{B}T \ll\,1$), we can derive an analytical expression for the chemical power for systems with small deviations from the mean thermodynamic equilibrium $\{\bar\phi_i^*\}$.
In this limit, the average chemical power in Eq.~\eqref{eq:chemical_power} takes a simpler form that depends on the linearised chemical force $\Delta \mu_c$ and reaction flux $\delta r$ around mean thermodynamic equilibrium as,
\begin{equation}
    \label{eq:power_linear}
    \mathcal{P}^c \simeq \Omega\int_0^{1/\Omega}\text{d}t'\left.\Delta\mu_c\right|_{\{\bar\phi_i^*\} }\dfrac{V^*}{\nu}\left.\delta r\right|_{\{\bar\phi_i^*\}}\,.
\end{equation}
Expanding chemical force upto linear order of deviations around $\{\bar\phi_i^*\}$ and using the expression for linearised reaction flux given  in Eq.~\eqref{eq:linear_chemical_flux}, the chemical power becomes
\begin{align}
\label{eq:power_lin_2}
    \mathcal{P}^c &=\frac{\Omega k_c^*V^*}{k_\text{B}T\nu}
\int_0^{1/\Omega}\text{d}t'\left[ \sum_{j\in\{A,B\}} \left(\mu_{Aj}^* - \mu_{Bj}^*\right)\delta\bar\phi_j(t') \right]^2
\end{align}
with $k_c^*=k_c\exp(\mu_A^*/k_\text{B}T) = k_c\exp(\mu_B^*/k_\text{B}T)$. 
The reservoir $x\in\{A,S\}$ pulls the system away from mean thermodynamic equilibrium, giving rise to deviations $\{\delta\bar\phi_i\}$ around $\{\bar\phi_i^*\}$. 
We introduce $m_i$ and $n_i$ as functions of $\widetilde\Omega$ as,
\begin{align}
\label{eq:mi_ni}
    m_i(\widetilde\Omega) &= \frac{\lambda_-}{\widetilde\Omega^2+\lambda_-^2}a_i + \frac{\lambda_+}{\widetilde\Omega^2+\lambda_+^2}b_i \, ,\\[0.5em]
    n_i(\widetilde\Omega) &=\frac{\widetilde\Omega}{\widetilde\Omega^2+\lambda_-^2}a_i + \frac{\widetilde\Omega}{\widetilde\Omega^2 + \lambda_+^2}b_i \, .
\end{align}
Using $m_i(\widetilde\Omega)$ and $n_i(\widetilde\Omega)$, we rewrite the expression for system deviations around mean thermodynamic equilibrium given in Eq.~\eqref{analytics:sol_5} as,
\begin{align}
    \delta\bar\phi_i(t) = \frac{\hat{\mu}^*_x}{k_\text{B}T}\left[m_i(\widetilde\Omega)\sin(\widetilde\Omega t) -n_i(\widetilde\Omega)\cos (\Omega t)\right]\,.
\end{align}
To obtain an analytical expression for power, we integrate the integral in Eq.~\eqref{eq:power_lin_2} of average chemical power over a period $t\,\in\,\left[0,1/\Omega\right]$. 
The expression of chemical power as a function of cycling frequency $\mathcal{P}^c(\Omega)$ as,
\begin{align}
\label{eq:analytical_power}    \mathcal{P}^c(\Omega) = \frac{\nu}{2V^*}\frac{k_c^*
    }{k_\text{B}T}\left(\frac{k_r^*\hat\mu_x}{k_\text{B}T}\right)^2\Bigg[\left(\mu_{AA}^* - \mu_{BA}^*\right)^2\left(m_A^2+n_A^2\right)\nonumber\\[0.5em]
    \left.
    +\left(\mu_{AB}^* - \mu_{BB}^* \right)^2\left(m_B^2 + n_B^2\right)\right.\nonumber\\[0.5em]
    +2\left(\mu_{AA}^* - \mu_{BA}^*\right)\left(\mu_{AB}^* - \mu_{BB}^*\right)
    \left(m_Am_B + n_An_B\right)\Bigg]\,.
\end{align}
with $k_r^* = k_r\exp(\left<\mu_x^r\right>/k_\text{B}T)=k_r\exp(\mu_x/k_\text{B}T)$ and $k_c^*=k_c\exp(\mu_A^*/k_\text{B}T) = k_c\exp(\mu_B^*/k_\text{B}T)$. We use the analytical expression of average chemical power from Eq.~\eqref{eq:analytical_power}, and plot power vs frequency for both solvent and reactant reservoir in Fig.~\ref{fig:amplitude_change}(c,d) (solid lines).

The analytical expression for chemical power in Eq.~\eqref{eq:analytical_power} is a function of cycling frequency $\Omega$.
In the limits of slow cycling frequencies ($\Omega/k_c\ll 1$), we find:
\begin{align}
\label{eq:lim_sf}
    \lim_{\Omega/k_c\rightarrow0} \,m_i(\Omega) &= \frac{a_i}{\lambda_-} + \frac{b_i}{\lambda_+}\,,
    \\
\lim_{\Omega/k_c\rightarrow0}\, n_i(\Omega)&=2\pi\Omega\left[\frac{a_i}{\lambda_-^2} + \frac{b_i}{\lambda_+^2}\right] \, . 
\end{align}
When substituting the above limits in average chemical power (Eq.~\eqref{eq:analytical_power}), we obtain
\begin{equation}  \mathcal{P}^c(\Omega) \propto \Omega^2  \, . 
\end{equation}

On the contrary, for fast cycling frequency ($\Omega/k_c\gg 1$), the functions $m_i$ and $n_i$ become:
\begin{align}
\label{eq:lim_ff}    \lim_{\Omega/k_c\rightarrow\infty} m_i(\Omega) &= 0\, ,\\
\lim_{\Omega/k_c\rightarrow\infty} n_i(\Omega)&=\frac{1}{2\pi\Omega}\left(a_i + b_i\right)\, .
\end{align}
In this limit, we see that the chemical power decays as follows:
\begin{equation}
    \mathcal{P}^c(\Omega)\propto\frac{1}{\Omega^2} \, . 
\end{equation}

\FloatBarrier
\clearpage
\onecolumngrid
\begin{table*}[t!] 
\centering
\caption{\textbf{Parameter choices for each figure:} Common for all figures in this work is that the molecular volumes are equal, $\nu_i=1$. All energy values in the table are given in units of $k_BT$. \label{tab:params}}
\setlength{\tabcolsep}{3.5pt}
\begin{tabular}{l c c c c c c c c c}
\hline
Figure & Reservoir &$\chi_{AS}/k_\text{B}T$ & $\chi_{BS}/k_\text{B}T$ & $\chi_{AB}/k_\text{B}T$ 
       & $\omega_A/k_\text{B}T$ & $\omega_B/k_\text{B}T$ & $\left<\mu^r_i/k_\text{B}T\right>$ & $\hat\mu_i/k_\text{B}T$  &comments\\
\hline
\ref{fig:orbit_shape_and_flux}(a,b,c)    & solvent & $3.0$  & $-0.6$  & $0.6$ & $0.4$ & $1.8$  &  $-0.35$ & $0.35$\\[0.7em]
\ref{fig:orbit_shape_and_flux}(d,e,f)    & reactant & $3.0$  & $-0.6$  & $0.6$ & $0.4$ & $1.8$   &  $-0.08$ & $0.35$\\[0.7em]
\ref{fig:amplitude_change}(a,b)    & reactant & $3.0$  & $-0.6$  & $0.6$ & $0.0$ & $0.05$   &  $-1.148$ & $0.6$\\[0.7em]
\ref{fig:dilute_system}    & reactant& $(0.0, 1.0)$  & $(0.0, 1.0)$  & $(0.0,1.0)$ & $0.0$ & $0.24$   &  $(-1.24, -0.84)$ & $0.49$ & (Ideal, Dilute) mixture\\[0.7em]
\ref{fig:dilute_system}    & solvent & $(0.0, 1.0)$  & $(0.0, 1.0)$  & $(0.0,1.0)$ & $0.0$ & $0.24$   &  $(-0.69, -0.50)$ & $0.49$ & (Ideal, Dilute) mixture\\[0.7em]
\ref{fig:diff_int_sys}(a)  &    & -0.6 & 3.0    & 2.5   & 1.8   & 1.0   &(0.691,-0.518) & 0.5   &(reactant, solvent) resv.\\[0.7em]
\ref{fig:diff_int_sys}(b) & &   0.6&    -0.6&   3.0 & 0.0   &0.24   & (-0.841,0.502)    & 0.49  & (reactant, solvent) resv. \\[0.7em]
\ref{fig:diff_int_sys}(c) & &   1.79&    -1.79 &   -1.79 & 0.4   &1.8   & (-0.327,-0.158)    & 0.158  & (reactant, solvent) resv. \\

\hline
\end{tabular}
\end{table*}
\twocolumngrid

%

\end{document}